\documentclass[12pt,reqno]{amsart}
\usepackage{sourceserifpro}
\usepackage{sourcecodepro}
\usepackage[T1]{fontenc}
\usepackage[italic,eulergreek,symbolmisc]{mathastext}
\usepackage{microtype}
\usepackage{natbib}
\usepackage[dvipsnames]{xcolor}
\usepackage{hyperref}
\usepackage{dsfont}
\usepackage{geometry}
\usepackage{setspace}
\usepackage{amssymb, color}
\usepackage{amsmath}
\usepackage{amsfonts}
\usepackage{rotating}
\usepackage[labelsep=none,format=default,labelfont={footnotesize},textfont={footnotesize},font={footnotesize},justification=justified,labelformat=empty]{caption}
\usepackage{appendix}
\usepackage{amssymb}
\usepackage{graphicx}
\allowdisplaybreaks

\MTsetmathskips{f}{\thinmuskip}{0mu}
\MTsetmathskips{y}{\thinmuskip}{0mu}
\MTsetmathskips{p}{\thinmuskip}{0mu}
\MTsetmathskips{l}{0mu}{\thinmuskip}
\MTsetmathskips{j}{\thinmuskip}{\thinmuskip}
\hypersetup{colorlinks=true,citecolor=Blue,filecolor=Blue,linkcolor=Blue,urlcolor=Blue}
\allowdisplaybreaks[4]

\def\E{\mathbb{E}}
\def\P{\mathbb{P}}
\def\R{\mathbb{R}}

\def\tr{\mathrm{tr}}
\def\Y{\mathrm{Y}}
\def\X{\mathrm{X}}
\def\e{\mathrm{e}}
\def\f{\mathrm{f}}
\def\hf{\widehat{\mathrm{f}}}
\def\I{\mathrm{I}}
\def\P{\mathrm{P}}
\def\M{\mathrm{M}}
\def\var{\mathrm{var}}

\begin{document}
\title[Tuning Parameter Selection in Econometrics]{Tuning Parameter Selection in Econometrics}
\author[Chetverikov]{Denis Chetverikov}
\address[D. Chetverikov]{Department of Economics, UCLA, Bunche Hall, 8283,
315 Portola Plaza, Los Angeles, CA 90095, USA.}
\email{chetverikov@econ.ucla.edu}
\date{\today }
\thanks{{\em Acknowledgements:} I thank Andrei Voronin and Lucas Zhang for excellent research assistance and Zhipeng Liao and Jesper Sorensen for useful discussions.}

\begin{abstract}
I review some of the main methods for selecting tuning parameters in nonparametric and $\ell_1$-penalized estimation. For the nonparametric estimation, I consider the methods of Mallows, Stein, Lepski, cross-validation, penalization, and aggregation in the context of series estimation. For the $\ell_1$-penalized estimation, I consider the methods based on the theory of self-normalized moderate deviations, bootstrap, Stein's unbiased risk estimation, and cross-validation in the context of Lasso estimation. I explain the intuition behind each of the methods and discuss their comparative advantages. I also give some extensions.
\end{abstract}

\maketitle

\section{Introduction}
Tuning parameter selection problem appears in many econometric settings. For example, one can think of selecting the number of terms in series estimation, selecting the bandwidth value in local estimation, selecting the penalty parameter in penalized estimation, selecting the number of factors in a factor model, selecting the number of components in a mixture model, and selecting the number of neurons and layers in a neural network. For all these settings, there exist multiple methods to select the corresponding tuning parameters. In this paper, I review some of them. For concreteness, I focus on the series estimation of nonparametric mean regression models and on $\ell_1$-penalized estimation of high-dimensional linear regression models. 

In the context of the series estimation, I consider the methods of Mallows, Stein, Lepski, cross-validation, penalization, and aggregation. The methods of Mallows, Stein, and cross-validation are based on the idea of unbiased risk estimation, which consists of selecting the tuning parameter, i.e. the number of series terms in our case, by minimizing an unbiased estimator of the risk of the series estimator. As I discuss, the advantage of the Mallows method is that it works not only for series estimators but in fact for any linear estimators. The Stein method requires the assumption of Gaussian noise but works also for non-linear estimators. Cross-validation is an almost universal method that can be applied in many settings well beyond the nonparametric mean regression model but may not be fully efficient ($V$-fold cross-validation) or may be hard to compute (leave-one-out cross-validation). The Lepski method is based on the idea of starting with a small number of series terms and increasing it up to a point where further increasing the number of terms does not yield a significant reduction of the bias of the estimator, where the significance is measured relative to the variance of the estimator. The penalization method is based on the idea of preventing estimation overfitting by penalizing estimators with too many series terms. Finally, instead of selecting the number of series terms, the aggregation method proposes using a weighted average of estimators with different numbers of series terms, with the idea that averaging the estimators may decrease their variance by averaging out the noise in individual estimators. The methods of Mallows, Stein, cross-validation, penalization, and aggregation are global in the sense that they are built to guarantee good average-over-the-domain performance of the series estimator, i.e. they come with estimation guarantees in the prediction and $\ell_2$ norms; see the next section for the definitions. In contrast, the Lepski method can be tuned to yield good performance at a particular point of interest in the domain. The latter is useful, for example, in regression discontinuity designs.

In the context of $\ell_1$-penalized estimation, I mainly focus on the Lasso estimator of the high-dimensional linear mean regression model. I consider methods to choose the penalty parameter for the Lasso estimator based on the theory of self-normalized moderate deviations, bootstrap, Stein's unbiased risk estimation, and cross-validation. The first two methods come with solid theoretical guarantees and the last two methods are intuitively attractive but require more theoretical work to fully understand their properties. I discuss existing results underlying each of these methods. In addition, I cover not only i.i.d. data setting but also clustered and panel data settings. Moreover, I explore extensions to quantile and generalized linear models.

It is surely impossible to describe all existing work on the problem of tuning parameter selection, and my review is in fact highly selective. In particular, I make no effort to trace the original contributions and, in many cases, give only few references which an interested reader can use as a starting point. Moreover, I omit some existing methods. For example, I do not discuss likelihood-based methods, such as AIC, BIC, and other Bayes-based methods. Also, I do not discuss plugin-type methods, where the tuning parameters are selected by a direct estimation of optimal population analogues, as those methods often lack coherent optimality properties. In addition, I do not cover methods suitable for time series data. Moreover, I focus exclusively on the tuning parameter selection problem from the estimation point of view and do not touch the problem from the inference point of view. I refer an interested reader to other available reviews, e.g. \cite{AC10}, \cite{DTY18}, \cite{WW20}, and \cite{BBL23}, which are all highly complementary to our discussion here.

\section{Tuning Parameter Selection for Nonparametric Estimation}
In this section, I discuss methods to choose tuning parameters in nonparametric settings. For concreteness, I introduce these methods in the context of the series estimator of the nonparametric mean regression models, but note that most methods to be discussed here can be adapted to other estimation problems as well. I consider the methods of Mallows, Stein, Lepski, cross-validation, penalization, and aggregation. At the end of this section, I also carry out a small Monte Carlo simulation study comparing finite-sample performance of these methods.

I consider the nonparametric mean regression model
\begin{equation}\label{eq: model}
Y = f(X) + e,\quad \E[e\mid X] = 0,
\end{equation}
where $Y\in\R$ is a dependent variable, $X\in\mathcal X$ is a vector of covariates, $e \in \R$ is a zero-mean noise, and $f\colon \mathcal X\to \R$ is a function of interest. I assume that our task is to estimate $f$ using a random sample $(X_1,Y_1),\dots,(X_n,Y_n)$ from the distribution of the pair $(X,Y)$. To describe the series estimator of the function $f$, for all $k\geq 1$, let $p^k = (p_1,\dots,p_k)^{\top}$ be a vector of $k$ basis functions mapping $\mathcal X$ into $\R$ and chosen so that a suitable linear combination of these functions can approximate $f$. Here, all $k$ functions in the vector $p^k$ are allowed to depend on $k$ but I omit this dependence for simplicity of notation. Assuming that $k$ is large enough, the approximation is expected to be sufficiently good in the sense that there exists a vector $\beta_k\in\R^k$ such that
\begin{equation}\label{eq: series approximation}
f(x)=p^k(x)^\top\beta_k + r_k(x), \quad x\in\mathcal X,
\end{equation}
where the residual $r_k$ is small. The series estimator is then based on the idea of OLS estimation of the vector of coefficients $\beta_k$ pretending that the residual $r_k$ is non-existent:
\begin{equation}\label{eq: series estimator}
\widehat f_k(\cdot) = p^k(\cdot)^\top \widehat \beta_k,  \quad\text{where}\quad \widehat \beta_k = \left(\sum_{i=1}^n p^k(X_i)p^k(X_i)^\top\right)^{-1}\sum_{i=1}^n p^k(X_i)Y_i.
\end{equation}
When $X$ is a scalar random variable, the basis functions could be monomials: $p_j(x)=x^{j-1}$ for $j=1,\dots,k$, or quadractic regression splines: $p_1(x)=1$, $p_2(x)=x$, $p_3(x)=x^2$, $p_j(x)=((x-(j-3)/(k-2))\vee 0)^2$ for $j=4,\dots,k$, if we further assume that the support of $X$ is the $[0,1]$ interval. Note that in the case of monomials, all functions $p_j$ are independent of $k$ but in the case of regression splines, only $p_1$, $p_2$, and $p_3$ are independent of $k$. When $X$ is a random vector, the basis functions could be given by tensor products of univariate monomials or regression splines, e.g. see \cite{BCCK15} for details and further examples of popular basis functions.

The series estimator $\widehat f_k$ in \eqref{eq: series estimator} contains a tuning parameter $k$, also known as a smoothing or regularization parameter, and one of the most important questions in the theory of nonparametric estimation is how this parameter can be chosen in practice. Intuitively, when $k$ is too small, the residual $r_k$ in \eqref{eq: series approximation} may not be negligible and the estimator $\widehat f_k$ may not be flexible enough to capture the shape of the function $f$. When $k$ is too large, we have to estimate too many parameters, which leads to the estimator $\widehat f_k$ with large variance. We thus want to choose $k$ in a way to strike a balance between flexibility of the estimator $\widehat f_k$ and its variance. Formally speaking, we would like to construct an estimator $\widehat k$ such that
\begin{equation}\label{eq: ideal result}
\frac{d(\widehat f_{\widehat k},f)}{\min_{k\in\mathcal K_n}d(\widehat f_k,f)}\to_P 1,
\end{equation}
where $d(\widehat f_{\widehat k},f)$ measures the distance between the estimator $\widehat f_{\widehat k}$ and the true function $f$, and $\mathcal K_n$ is a set of candidate values of the tuning parameter $k$ that is chosen large enough to include a value striking a good balance between flexibility and variance. Here, the set $\mathcal K_n$ is typically assumed to depend on $n$ and to grow together with $n$. For example, when the basis functions are given by monomials, we could consider $\mathcal K_n = \{1,\dots,\bar k_n\}$, where $\bar k_n = [n^{1/3}]$, the largest integer smaller than $n^{1/3}$. When the basis functions are given by quadratic splines, we could consider $\mathcal K_n = \{3,\dots,\bar k_n\}$, where $\bar k_n = [n^{1/2}]$, the largest integer smaller than $n^{1/2}$.\footnote{As discussed in \cite{BCCK15}, $\bar k_n$ should be chosen so that $\bar k_n^2\log n = o(n)$ in the case of monomials and $\bar k_n\log n = o(n)$ in the case of quadratic splines in order to guarantee existence of the series estimator. In addition, to avoid numerical problems with inverting the matrix $\sum_{i=1}^n p^k(X_i)p^k(X_i)^{\top}$, it is recommended to consider a suitable linear transformation of the functions $p^k$. For example, one can use Legendre polynomials instead of monomials and B-splines instead of regression splines.}

The following four distance measures are typically considered in the literature:
\begin{align}
\text{$\ell_2$ metric: } & d(g,f) = \sqrt{\E[(g(X) - f(X))^2]}, \label{eq: l2 metric}\\
\text{prediction metric: } & d(g,f) = \sqrt{\textstyle{n^{-1}\sum_{i=1}^n} (g(X_i) - f(X_i))^2}, \label{eq: prediction metric}\\
\text{uniform metric: } & d(g,f) = \sup|g(x) - f(x)|, \label{eq: linf metric}\\
\text{pointwise metric: } & d(g,f) = |g(x_0) - f(x_0)|, \label{eq: lpw metric}
\end{align}
where the supremum in the uniform metric is taken either over all $x$ in $\mathcal X$ or over all $x$ in some user-specified subset $\bar{\mathcal X}$ of $\mathcal X$, and $x_0$ in the pointwise metric is a user-specified value in $\mathcal X$. A result in the $\ell_2$ metric is typically useful in the semi-parametric/double-machine-learning estimation, where the estimator $\widehat f_{\widehat k}$ is used as a first step in a two-step procedure to learn some causal parameters, e.g. see \cite{CCDDHNR18}. A result in the prediction metric is useful in signal processing, where the tast is to get rid of the noise $e_i$ in observable signals $Y_i$. A result in the uniform metric is useful when we want to understand the shape of the function $f$. A result in the pointwise metric is useful when we are interested in a particular point $x_0$ in the domain of the function $f$. Note that the results of the form \eqref{eq: ideal result} in the $\ell_2$ and prediction metrics are closely related, as it is typically possible to translate one into the other by applying the law of large numbers.

When the estimator $\widehat f_{\widehat k}$ satisfies \eqref{eq: ideal result}, it is said to be {\em asymptotically optimal}. However, the result of the form \eqref{eq: ideal result} may not always be achievable. Intuitively, a problem occurs when there exists $k\in\mathcal K$ such that $d(\widehat f_k,f)$ is too small, which is the case when a linear combination of the functions $p_1,\dots,p_k$ yields the function $f$ exactly. Alternatively, the result \eqref{eq: ideal result} may be achievable but the rate of convergence to the constant one may be slow. For these reasons, we are also interested in constructing an estimator $\widehat k$ such that for some constant $C\geq 1$ and a sequence $\varepsilon_n$ converging to zero, we have either
\begin{equation}\label{eq: oracle probability}
d(\widehat f_{\widehat k},f) \leq C\min_{k\in\mathcal K_n}d(\widehat f_k,f) + \varepsilon_n
\end{equation}
with probability approaching one or
\begin{equation}\label{eq: oracle expectation}
\E[d(\widehat f_{\widehat k},f)] \leq C\min_{k\in\mathcal K_n}\E\left[d(\widehat f_k,f)\right] + \varepsilon_n.
\end{equation}
Such inequalities are called {\em oracle inequalities} because they show the performance of the estimator $\widehat f_{\widehat k}$ based on our choice of $k$ relative to the performance of the estimator $\widehat f_{k}$ based on the oracle choice $k = k^O$, where 
$$
k^O = \arg\min_{k\in\mathcal K_n} d(\widehat f_k,f),
$$
The purpose of this section is to discuss estimators $\widehat k$ that lead to results in the forms \eqref{eq: ideal result}, \eqref{eq: oracle probability}, \eqref{eq: oracle expectation}, or in some related forms.


For our discussion in this section, it will be convenient to use matrix notations. I therefore denote $\Y = (Y_1,\dots,Y_n)^{\top}$, $\f = (f(X_1),\dots,f(X_n))^{\top}$, and $\e = (e_1,\dots,e_n)^{\top}$, so that $\Y = \f + \e$. Also, I denote $\hf_k = (\widehat f_k(X_1),\dots,\widehat f_k(X_n))^{\top}$, $\P_k = (p^k(X_1),\dots,p^k(X_n))^{\top}$, and $\M_k = \P_k(\P_k^{\top}\P_k)^{-1}\P_k^{\top}$, so that $\hf_k  = \M_k \Y$. In addition, I let $\I_k$ be a $k\times k$ identity matrix. Moreover, I denote $\sigma^2_i = \E[e_i^2\mid X_i]$ for all $i=1,\dots,n$ and let $\Omega$ be the $n\times n$ diagonal matrix with $(\sigma_1^2,\dots,\sigma_n^2)$ on the diagonal. Finally, for any function $g\colon \mathcal X\to\R$, I denote $\|g\|_{2,n}=(n^{-1}\sum_{i=1}^n g(X_i)^2)^{1/2}$.

\subsection{The Method of Mallows}\label{sub: mallows} \cite{M73} proposed a method that in our context of choosing $k$ for the series estimator of model \eqref{eq: model} takes the following form:
\begin{equation}\label{eq: mallows definition}
\widehat k^M = \arg\min_{k\in\mathcal K}\left\{ \frac{1}{n}\sum_{i=1}^n (Y_i - \widehat f_k(X_i))^2 + \frac{2}{n}\tr(\M_k\Omega) - \frac{1}{n}\tr(\Omega) \right\}.
\end{equation}
To explain the intuition behind this method, observe that since $\hf_k = \M_k\Y = \M_k \f + \M_k \e$, we have
$$
\E[\|\hf_k - \f\|^2] = \E[\| (\M_k - \I_k)\f + \M_k \e \|^2] = \E[\| (\M_k - \I_k)\f \|^2] + \E[\|\M_k \e \|^2].
$$
Also, given that $\Y = \f + \e$ and $\hf = \M_k \f + \M_k \e$, we have
\begin{align*}
\E[\| \Y - \hf \|^2] 
&= \E[\| (\I_k - \M_k)\f + (\I_k - \M_k)\e \|^2] \\
&=\E[\| (\I_k - \M_k)\f \|^2] + \E[\|(\I_k - \M_k)\e \|^2].
\end{align*}
Here,
\begin{align*}
\E[\|(\I_k - M_k) \e\|^2] 
&= \E[\e^{\top}(\I_k - \M_k)^{\top}(\I_k - \M_k) \e] \\
& =\E[\|\e\|^2]  - 2\E[\e^{\top}\M_k\e] + \E[\|\M_k\e\|^2].
\end{align*}
Moreover, $\E[\|\e\|^2] = \E[\tr(\Omega)]$ and
$$
\E[\e^{\top}\M_k \e] = \E[\tr(\e^{\top}\M_k \e)] = \E[\tr(\M_k \e\e^{\top})] = \E[\tr(\M_k\Omega_k)]
$$
by interchanging the trace and the expectation operators and using the law of iterated expectations. Thus,
\begin{equation}\label{eq: mallows is unbiased}
\E\left[ \frac{1}{n}\sum_{i=1}^n (Y_i - \widehat f_k(X_i))^2 + \frac{2}{n}\tr(\M_k\Omega) - \frac{1}{n}\tr(\Omega) \right] = \E\left[\| \widehat f_k - f \|^2_{2,n}\right].
\end{equation}
The left-hand side here is the expectation of the criterion function in \eqref{eq: mallows definition} and the right-hand side is often referred to as the risk of the estimator $\widehat f_k$, which we want to minimize. The Mallows method \eqref{eq: mallows definition} thus consists of minimizing an unbiased risk estimator.
In addition, unless the set $\mathcal K_n$ is too large, standard concentration inequalities, e.g. \cite{BLM13}, can be used to show that the random variables
$$
\frac{1}{n}\sum_{i=1}^n (Y_i - \widehat f_k(X_i))^2 + \frac{2}{n}\tr(\M_k\Omega) - \frac{1}{n}\tr(\Omega)\quad\text{and}\quad \| \widehat f_k - f \|^2_{2,n}
$$
converge in probability to their expectations uniformly over $k$. Convergence of the former random variable, together with \eqref{eq: mallows definition} and \eqref{eq: mallows is unbiased}, implies that $\widehat k^M$ nearly minimizes the risk $\E[\| \widehat f_k - f \|^2_{2,n}]$. Convergence of the latter random variable implies that $\widehat k^M$ also nearly minimizes $\| \widehat f_k - f \|^2_{2,n}$. Formally, under appropriate regularity conditions, we can then obtain the following result:
\begin{equation}\label{eq: li result}
\frac{\| \widehat f_{\widehat k^M} - f \|^2_{2,n}}{\min_{k\in\mathcal K_n}\| \widehat f_k - f \|^2_{2,n}}\to_P 1;
\end{equation}
see \cite{L87} and \cite{A91} for details in the homoskedastic and heteroskedastic cases, respectively. Thus, the Mallows method yields an estimator that is asymptotically optimal in the prediction metric. Moreover, as discussed above, \eqref{eq: li result} can be combined with the uniform law of large numbers to obtain asymptotic optimality in the $\ell_2$ metric. Related results were also obtained in \cite{Shi81, PT90, K94}.

Note, however, that the appropriate regularity conditions here include the requirement that $n\|\widehat f_k - f\|_{2,n}^2\to \infty$ for all $k\in\mathcal K_n$, and this requirement is not satisfied when $f(\cdot) = p^k(\cdot)^{\top}\beta_k$ for some $k\in\mathcal K_n$ and $\beta_k \in \R^k$, as follows from the standard OLS estimation theory. To obtain a more general result for the series estimator $\widehat f_k$ with the number of series terms $k$ selected by the Mallows method, we thus need to proceed through oracle inequalities. We will return to this topic in Section \ref{sec: penalization method}, where we discuss the penalization method.


The Mallows method, as described above, is generally infeasible as it requires the knowledge of the heteroskedasticity matrix $\Omega$. However, it is fairly straightforward to obtain a feasible version of this method. Indeed, observe that the term $n^{-1}\tr(\Omega)$ in the criterion function in \eqref{eq: mallows definition} is independent of $k$ and thus can be omitted. In addition, given that $\M_k=\P_k(\P_k^{\top}\P_k)^{-1}\P_k$, we have
\begin{equation}\label{eq: trace trick}
\tr(\M_k\Omega) = \tr((\P_k^{\top}\P_k)^{-1}\P_k^{\top}\Omega\P_k),
\end{equation}
which can be consistently estimated by
\begin{equation}\label{eq: ak def}
\widehat A_k = \tr\left\{ \left(\sum_{i=1}^n p^k(X_i)p^k(X_i)^{\top}\right)^{-1}\sum_{i=1}^n \widehat e_i^2 p^k(X_i)p^k(X_i)^{\top} \right\},
\end{equation}
where $\widehat e_i = Y_i - \widehat f_{\bar k}(X_i)$ for all $i=1,\dots,n$ and some $\bar k\in\mathcal K_n$ that is known to yield a consistent estimator $\widehat f_{\bar k}$ of the function $f$, e.g. $\bar k = \bar k_n = [n^{1/3}]$. A feasible version of the Mallows method is thus given by
\begin{equation}\label{eq: mallows feasible version}
\widehat k^M = \arg\min_{k\in\mathcal K_n}\left\{\frac{1}{n}\sum_{i=1}^n (Y_i - \widehat f_k(X_i))^2 + \frac{2\widehat A_k}{n}\right\}.
\end{equation}
Moreover, note that if we have a model with homoskedastic noise, i.e. $\sigma_i^2 = \sigma^2$ for all $i=1,\dots,n$ and some $\sigma^2>0$, the Mallows estimator $\widehat k^M$ in \eqref{eq: mallows definition} reduces to
\begin{equation}\label{eq: mallows method homoskedasticity}
\widehat k = \arg\min_{k\in\mathcal K_n}\left\{\frac{1}{n}\sum_{i=1}^n (Y_i - \widehat f_k(X_i))^2 + \frac{2\sigma^2 k}{n}\right\},
\end{equation}
as the expression in \eqref{eq: trace trick} in this case reduces to $\tr((\P_k^{\top}\P_k)^{-1}(\P_k^{\top}\P_k)) = \tr(\I_k) = k$. We will use this formula later to relate the Mallows and penalization methods.

Finally, note that the intuition behind the estimator $\widehat k^M$ in \eqref{eq: mallows definition} does not really depend on the properties of the series estimator $\widehat \f_k = \M_k\Y$ except for its linearity in $\Y$, and so the Mallows method can in fact be used to choose the regularization parameter $k$ for other linear in $\Y$ estimators of the function $f$, e.g. ridge estimators. In addition, the Mallows method is often used for aggregating linear estimators, i.e. for constructing estimators of the form $\sum_{k\in\mathcal K_n} w_k\widehat f_k$, where the weights $w_k$ are selected by solving an optimization problem analogous to that in \eqref{eq: mallows definition}, e.g. see \cite{H07} in the context of the series estimator and \cite{ZWZZ19} and \cite{ZWZL23} in some related contexts.

\subsection{The Method of Stein}\label{sub: stein}
\cite{S81} proposed the following estimator of $k$:
\begin{equation}\label{eq: stein definition}
\widehat k^S = \arg\min_{k\in\mathcal K}\left\{\frac{1}{n}\sum_{i=1}^n (Y_i - \widehat f_k(X_i))^2 + \frac{2}{n}\tr(D^k\Omega) - \frac{1}{n}\tr(\Omega)\right\},
\end{equation}
where $D^k$ is the $n\times n$ matrix defined by $D^k_{ij} = \partial\widehat f_k(X_i)/\partial Y_j$ for all $i,j=1,\dots,n$. This method is only justified when the conditional distribution of $e$ given $X$ is Gaussian, which we are going to assume throughout this section. To see the intuition behind this method, observe that if $Z\sim N(\mu,\sigma^2)$ and $g$ is a (weakly) differentiable function, then $\sigma^2\E[g'(Z)] = \E[(Z-\mu)g(Z)]$ via integration by parts. Applying this identity, which is often referred to as Stein's lemma, conditional on $\mathrm X = (X_1,\dots,X_n)$ to the function $Y_i\mapsto \widehat f_k(X_i) - Y_i$ yields
$$
\sigma_i^2\E\left[\frac{\partial}{\partial Y_i}(\widehat f_k(X_i) - Y_i)\mid \mathrm X\right] = \E[(Y_i - f(X_i))(\widehat f_k(X_i) - Y_i)\mid \mathrm X],\quad \text{for all }i=1,\dots,n.
$$
Thus,
\begin{align*}
& \E\left[ \frac{1}{n}\sum_{i=1}^n (Y_i - \widehat f_k(X_i))^2 + \frac{2}{n}\tr(D^k\Omega) - \frac{1}{n}\tr(\Omega) \right] \\
& \qquad = \E\left[ \frac{1}{n}\sum_{i=1}^n (Y_i - \widehat f_k(X_i))^2 + \frac{2}{n}\sum_{i=1}^n(Y_i - f(X_i))(\widehat f_k(X_i) - Y_i) + \frac{1}{n}\sum_{i=1}^n(Y_i - f(X_i))^2 \right] \\
& \qquad = \E\left[ \frac{1}{n}\sum_{i=1}^n(\widehat f_k(X_i) - f(X_i))^2 \right] = \E\left[\|\widehat f_k - f\|_{2,n}^2\right].
\end{align*}
Hence, like the the Mallows method, the Stein method minimizes an unbiased risk estimator. Using the same arguments as those used to analyze the Mallows method, it is thus possible to show that the Stein method yields an asymptotically optimal estimator $\widehat f_{\widehat k^S}$ in the prediction and $\ell_2$ metrics. In the literature, this method is often referred to as the Stein's unbiased risk estimation, or SURE, for short.

The advantage of the Stein method over the Mallows method is that it applies for non-linear in $\Y$ estimators as well. In fact, it only requires that the estimators $\widehat f_k$ are weakly differentiable in $\Y$ and can be used to choose regularization parameters of estimators as complicated as the Lasso estimator, which we are going to discuss in Section \ref{sec: Stein lasso} below. In addition, for example, \cite{MW00} applied the Stein method to estimation problems with shape restrictions and \cite{CST13} applied it to low-rank matrix recovery problems. On the other hand, the disadvantage of the Stein method is that it requires that the conditional distribution of $e$ given $X$ is Gaussian, and it seems generally difficult to relax this condition, unless the estimators $\widehat f_k$ are sufficiently smooth in $\Y$.

It is also interesting to note that in our special case of the series estimator, the Stein method yields $\widehat k^S$ that is identical to $\widehat k^M$ given by the Mallows method. Indeed, given that
$$
\widehat f_k(X_i) = p^k(X_i)^{\top}\left(\sum_{j=1}^n p^k(X_j)p^k(X_j)^{\top}\right)^{-1}\sum_{j=1}^n p^k(X_j)Y_j,\quad\text{for all }i=1,\dots,n,
$$
we have
$$
D_{ii}^k = p^k(X_i)^{\top}\left(\sum_{j=1}^n p^k(X_j)p^k(X_j)^{\top}\right)^{-1}p^k(X_i),\quad \text{for all }i=1,\dots,n,
$$
and so $\tr(D^k\Omega) = \tr(\M_k\Omega)$, meaning that the criterion functions in \eqref{eq: mallows definition} and \eqref{eq: stein definition} coincide. In particular, this implies that a feasible version of the Stein method can be obtained as discussed in the case of the Mallows method.

\subsection{The Method of Lepski}
\cite{L90, L91, L92} proposed a method for choosing regularization parameters for nonparametric estimators that has an advantage of yielding results not only in the prediction and $\ell_2$ metrics but also in the uniform and pointwise metrics. The method attracted substantial attention in the literature and many variants of the method have been developed. Here, we describe a slight modification of the variant proposed in \cite{SW19}. For concreteness, I focus on results in the pointwise metric.

Suppose that we are interested in estimating $f(x_0)$ for some particular $x_0\in\mathcal X$ and our task is to choose one estimator from a collection of the series estimators $\{\widehat f_k(x_0)\}_{k\in\mathcal K_n}$. Intuitively, as we increase the number of series terms $k$, the (absolute value of the) bias $|\E[\widehat f_k(x_0) - f(x_0)]|$ of the estimator $\widehat f_k(x_0)$ goes down and its variance $\var(\widehat f_k(x_0))$ goes up. The main idea of the Lepski method is to start with a small value of $k$ and increase it up to a point where further increasing it does not significantly decrease the bias of the estimator $\widehat f_k(x_0)$, where the significance is understood relative to the variance of the estimator. To formalize this intuition, for all $k,k'\in\mathcal K_n$, denote
$$
b_{k,k'} = \E[\widehat f_{k'}(x_0) - \widehat f_{k}(x_0)]\quad\text{and}\quad p_{k,k'} = \var(\widehat f_{k'}(x_0) - \widehat f_{k}(x_0)).
$$
Also, let $\beta>0$ be some number and let $\mathcal K^+_n(k) = \{k'\in\mathcal K_n\colon k'>k\}$. Then for each $k\in\mathcal K_n$, we can consider the problem of testing the null hypothesis
$$
\mathrm H_{k}\colon |b_{k,k'}|^2 \leq \beta^2p_{k,k'},\quad\text{for all }k'\in\mathcal K_n^+(k).
$$
If the hypothesis $\mathrm H_{k}$ is not rejected, there is no reason to set the number of series terms above $k$. The method thus consists of choosing the number of series terms as the smallest number $k\in\mathcal K_n$ such that $\mathrm H_k$ is not rejected.\footnote{The original Lepski method considered choosing the number of series terms as the smallest number $k\in\mathcal K_n$ such that $\mathrm H_{k'}$ for all $k'\geq k$ are not rejected. The more relaxed choice described here was proposed in \cite{SV09}.}

It thus remains to discuss appropriate tests for hypotheses $\mathrm H_{k}$. One option is as follows. Denote $T_{k,k'}=|\widehat f_{k'}(x_0) - \widehat f_{k}(x_0)|$ for all $k'\in\mathcal K_n^+(k)$. Then under the null hypothesis $\mathrm H_{k}$, we have for all $k'\in\mathcal K_n^+(k)$ that $|b_{k,k'}|\leq \beta \sqrt{p_{k,k'}}$ and so
\begin{align*}
\frac{T_{k,k'}}{\sqrt{p_{k,k'}}} 
& \leq \beta + \frac{|\widehat f_{k'}(x_0) - \E[\widehat f_{k'}(x_0)] - \widehat f_{k}(x_0) + \E[\widehat f_{k}(x_0)]|}{\sqrt{p_{k,k'}}}.
\end{align*}
Thus, letting $c_{k}(\alpha)$ be the $(1-\alpha)$ quantile of
\begin{equation}\label{eq: lepski key random variable}
\max_{k'\in\mathcal K_n^+(k)}\frac{|\widehat f_{k'}(x_0) - \E[\widehat f_{k'}(x_0)] - \widehat f_{k}(x_0) + \E[\widehat f_{k}(x_0)]|}{\sqrt{p_{k,k'}}},
\end{equation}
a test of $\mathrm H_{k}$ with size $\alpha$ consists of rejecting $\mathrm H_{k}$ if
$$
\max_{k'\in\mathcal K_n^+(k)} \frac{T_{k,k'}}{\sqrt{p_{k,k'}}} > \beta + c_{k}(\alpha).
$$
Let $\widehat k^L$ be the value of $k\in\mathcal K_n$ selected by this method.

To understand properties of the estimator $\widehat f_{\widehat k^L}(x_0)$, let $k^*$ be the smallest value of $k\in\mathcal K_n$ such that the hypothesis $\mathrm H_k$ is true. We can think of $\widehat f_{k^*}(x_0)$ as the oracle estimator, and our main task is to derive a bound on how much the estimator $\widehat f_{\widehat k^L}(x_0)$ deviates from this oracle estimator.\footnote{The name ``oracle estimator'' comes from the observation that we can think of $\widehat f_{k^*}(x_0)$ as the estimator provided by the oracle, who does not need to test hypotheses $\mathrm H_k$ as it knows whether these hypotheses are true or not.} To do so, observe that by construction,  the hypothesis $\mathrm H_{k^*}$ is accepted with probability at least $1-\alpha$, and so the probability that $\widehat k^L\leq k^*$ is also at least $1 - \alpha$. Moreover, letting $\mathcal K_n^-(k^*) = \{k\in\mathcal K_n\colon k<k^*\}$, we have for any $k\in \mathcal K_n^-(k^*)$ that
\begin{align*}
\frac{T_{k,k^*}}{\sqrt{p_{k,k^*}}} 
&\geq \frac{|b_{k,k^*}|}{\sqrt{p_{k,k^*}}} - \frac{|\widehat f_{k}(x_0) - \E[\widehat f_{k}(x_0)] - \widehat f_{k^*}(x_0) + \E[\widehat f_{k^*}(x_0)]|}{\sqrt{p_{k,k^*}}}.
\end{align*}
Hence, letting $c_{k^*}^-(\alpha)$ be the $(1-\alpha)$ quantile of
$$
\max_{k\in\mathcal K_n^-(k^*)}\frac{|\widehat f_{k}(x_0) - \E[\widehat f_{k}(x_0)] - \widehat f_{k^*}(x_0) + \E[\widehat f_{k^*}(x_0)]|}{\sqrt{p_{k,k^*}}}
$$
and 
$$
\mathcal K_n^0 = \left\{k\in \mathcal K_n^-(k^*)\colon |b_{k,k^*}|^2 > p_{k,k^*}(c_k(\alpha) + c_{k^*}^-(\alpha))^2\right\},
$$
we have
$$
\frac{T_{k,k^*}}{\sqrt{p_{k,k^*}}} > c_{k}(\alpha),\quad\text{for all }k\in\mathcal K_n^0,
$$
with probability at least $1-\alpha$. Thus, by the union bound, $\widehat k^L\in \mathcal K_{IN}$ with probability at least $1 - 2\alpha$, where $\mathcal K_{IN} = \{k\in\mathcal K_n\colon k\leq k^*\}\setminus \mathcal K_n^0$. In turn,
\begin{equation}\label{eq: lepski bound}
|\widehat f_{\widehat k^L}(x_0) - \widehat f_{k^*}(x_0)| \leq \max_{k\in\mathcal K_{IN}} \sqrt{p_{k,k^*}}(\beta + c_k(\alpha))
\end{equation}
with the same probability.

The quantity on the right-hand side of \eqref{eq: lepski bound} shows how much the series estimator $\widehat f_{\widehat k^L}(x_0)$ with the number of series terms determined by the Lepski method deviates from the oracle estimator $\widehat f_{k^*}(x_0)$ and can be understood as the adaptation price, as it shows how much we need to pay in terms of the precision of the estimator for not knowing the oracle choice $k^*$. To understand this quantity, observe first that $c_k(\alpha)$ is typically of order $\sqrt{\log|\mathcal K_n|} \leq \sqrt{\log n}$, where $|\mathcal K_n|$ denotes the number of elements in the set $\mathcal K_n$. Also, we typically have that $p_{k,k^*}$ is of order $(k^*-k)/n$. Thus, the size of the adaptation price depends on the set $\mathcal K_{IN}$, which \cite{SW19} refer to as the {\em insensitivity region}. If the function $k\mapsto b_{k,k^*}$ is relatively flat around $k=k^*$, this region may be rather large, in which case we expect the adaptation price to be of order $\sqrt{k^*\log n/n}$. Since the oracle estimation error $\widehat f_{k^*}(x_0) - f(x_0)$ is typically of order $\sqrt{k^*/n}$, it follows that the adaptation price in this case is actually larger than the oracle estimation error itself. On the other hand, if the function $k\mapsto b_{k,k^*}$ is relatively steep around $k=k^*$, the insensitivity region may be rather small, in which case the adaptation price may be as small as $\sqrt{\log n/n}$ in order, and may be much smaller than the oracle estimation error.

Next, observe that implementing the Lepski method requires the choice of two parameters, $\alpha$ and $\beta$, and one could potentially argue that the Lepski method is strange in the sense that it replaces the problem of choosing one tuning parameter, $k$, by the problem of choosing two tuning parameters, $\alpha$ and $\beta$. However, these parameters are very different from $k$ as they do not generate the bias-variance trade-off. Instead, these parameters specify the properties of the method. Indeed, the purpose of $\beta$ is to fix the oracle estimator we are potentially interested in, and the choice $\beta = 1$ seems reasonable. Similarly, the purpose of $\alpha$ is to specify the probability for theoretical guarantees, and is closely related to the purpose of the size control parameter in testing problems. The choices $\alpha = 0.05$ and $\alpha = 0.01$ seem reasonable.

The Lepski method as described here yields results in the pointwise metric. However, it is fairly straightforward to extend the method to yield results in other metrics as well, which is its main advantage. Indeed, if we want to have results in the uniform metric, we simply need to replace the statistic $T_{k,k'} = |\widehat f_{k'}(x_0) - \widehat f_k(x_0)|$ by $T_{k,k'} = \sup_{x\in\bar{\mathcal X}}|\widehat f_{k'}(x) - \widehat f_{k}(x)|$, and if we want the results in the $\ell_2$ metric, we should use $T_{k,k'} = (\E_X[(\widehat f_{k'}(X) - \widehat f_k(X))^2])^{1/2}$ or its estimated version, where the expectation is taken with respect to the distribution of $X$ that is independent of the data. Of course, in both case, the critical values for the tests should be adjusted accordingly. Note that other metrics are possible here as well. On the other hand, a disadvantage of the method is that it yields results relative to the oracle number of series terms $k^*$ rather than relative to the more relevant oracle number $k^{O} = \arg\min_{k\in\mathcal K}|\widehat f_k(x_0) - f(x_0)|$ appearing above.

Further, the variant of the method described so far is infeasible as it requires the knowledge of $p_{k,k'}$'s and $c_k(\alpha)$'s. However, these quantities are easy to estimate. Indeed, denoting $\mathrm Q_k = n^{-1}\sum_{i=1}^n p^k(X_i)p^k(X_i)^{\top}$ for all $k\in\mathcal K_n$ and letting $\bar k$ be some value of $k\in\mathcal K_n$ that is known to yield a consistent estimator $\widehat f_{\bar k}$ of the function $f$, e.g. $\bar k = \bar k_n = [n^{1/3}]$, and denoting $\widehat e_i = Y_i - \widehat f_{\bar k}(X_i)$ for all $i=1,\dots,n$, it follows that we can estimate $p_{k,k'}$ by
$$
\widehat p_{k,k'} = \frac{1}{n^2}\sum_{i=1}^n \widehat e_i^2 \Big(p^{k'}(x_0)^{\top}\mathrm Q_{k'}^{-1}p^{k'}(X_i) - p^{k}(x_0)^{\top}\mathrm Q_{k}^{-1}p^{k}(X_i)\Big)^2
$$
for all $k,k'\in\mathcal K_n$. Also, $c_k(\alpha)$ can be estimated, for example, by the multiplier bootstrap: let $\{w_i\}_{i=1}^n$ be i.i.d. standard Gaussian random variables and let
\begin{equation}\label{eq: lepski bootstrap}
\frac{n^{-1}\sum_{i=1}^n(p^{k'}(x_0)^{\top}\mathrm Q_{k'}^{-1} p^{k'}(X_i) - p^{k}(x_0)^{\top}\mathrm Q_{k}^{-1} p^{k}(X_i))w_i\widehat e_i}{\sqrt{\widehat p_{k,k'}}}
\end{equation}
be the bootstrap version of the random variable in \eqref{eq: lepski key random variable}. Then $c_k(\alpha)$ can be estimated by $\widehat c_k(\alpha)$, the $(1-\alpha)$ quantile of the conditional distribution of the random variable in \eqref{eq: lepski bootstrap} given the data. Consistency of (the properly normalized versions of) the estimators $\widehat p_{k,k'}$ and $\widehat c_{\alpha}(k)$ can be established using the results in \cite{BCCK15} and \cite{CCK13, CCK17, CCKK22}.

Finally, I note that the Lepski method is nearly universal in the sense that it can be adjusted to work in other estimation frameworks as long as the estimators in these frameworks feature a bias-variance trade-off and can be ordered in terms of decreasing bias. For example, \cite{CC18} and \cite{CCKan21} used this method in the context of nonparametric instrumental variable estimation, \cite{SWH13} used it in the context of nonparametric quantile regression, and \cite{CCK14}, among many others, used it in the context of density estimation.

\subsection{The Method of Cross-Validation} Cross-validation, which is traditionally attributed to the work of \cite{A74}, \cite{S74}, and \cite{G75}, is one of the most universal and popular methods to choose tuning parameters in general and in the context of mean regression models via the series estimator in particular. Below, we will consider its three main versions: validation, $V$-fold cross-validation, and leave-one-out cross-validation.

To describe the validation method, let $(I_1,I_2)$ be a random split of the whole sample $I_0 = \{1,\dots,n\}$ into two subsamples, so that $I_1\cap I_2 = \emptyset$ and $I_1\cup I_2 = I_0$. Here, $I_1$ and $I_2$ are referred to as the training subsample and the validation subsample, respectively. Typically, $I_1$ and $I_2$ are assumed to contain roughly two thirds and one third of the whole sample, respectively, but this will not be important for our discussion. For each $k$ in the set of candidate values $\mathcal K_n$, we build a series estimator of the function $f$ using the training subsample only: $\widehat f_{k,1}(\cdot) = p^k(\cdot)^{\top}\widehat\beta_k$, where
$$
\widehat\beta_k = \left(\sum_{i\in I_1}p^k(X_i)p^k(X_i)^{\top}\right)\sum_{i\in I_1}p^k(X_i)Y_i.
$$
We then choose $k$ by minimizing the out-of-sample prediction error of these estimators:
\begin{equation}\label{eq: validation definition}
\widehat k^{V} = \arg\min_{k\in\mathcal K_n} \sum_{i\in I_2}(Y_i - \widehat f_{k,1}(X_i))^2.
\end{equation}
The ultimate estimator of the function $f$ is then typically $\widehat f_V(\cdot) = \widehat f_{\widehat k^V,1}(\cdot)$.

Theoretical properties of the validation procedure are studied, among others, in \cite{W03} and \cite{M07}. Intuitively, conditional on the data in the training subsample $\{(X_i,Y_i\}_{i\in I_1}$, the sum $\sum_{i\in I_2}(Y_i - \widehat f_{k,1}(X_i))^2$ consists of i.i.d. terms with mean 
$$
\E_{X,Y}[(Y - \widehat f_{k,1}(X))^2] = \E_{X,Y}[e^2] + \E_{X,Y}[(\widehat f_{k,1}(X) - f(X))^2],
$$
where $\E_{X,Y}[\cdot]$ denotes the expectation with respect to the distribution of the pair $(X,Y)$ holding the data $\{(X_i,Y_i\}_{i\in I_1}$, and thus $\widehat f_{k,1}$, fixed. Thus, $\widehat k^V$ in \eqref{eq: validation definition} approximately minimizes $\E_{X,Y}[(\widehat f_{k,1}(X) - f(X))^2]$ and can be shown to lead to a result in the form of \eqref{eq: ideal result} in the $\ell_2$ metric with $\widehat f_k$ replaced by $\widehat f_{k,1}$.

Next, although the validation procedure described above is very intuitive, an issue with this procedure is that it yields the result for the subsample estimators $\widehat f_{k,1}$ instead of the full-sample estimators $\widehat f_k$, which are inefficient because they do not use the validation subsample in estimation. To fix this issue, we can consider $V$-fold cross-validation. To describe it, let $(I_1,\dots,I_V)$ be a random partition of the whole sample $I_0 = \{1,\dots,n\}$ into $V$ subsamples of roughly the same size. For each $v=1,\dots,V$, we build a series estimator of the function $f$ using all the data excluding observations in subsample $v$: $\widehat f_{k,v}(\cdot) = p^k(\cdot)^{\top}\widehat\beta_{k,v}$, where
$$
\widehat\beta_{k,v} = \left(\sum_{i\in I_0\setminus I_v} p^k(X_i)p^k(X_i)^{\top}\right)^{-1}\sum_{i\in I_0\setminus I_v}p^k(X_i)Y_i.
$$
We then choose the tuning parameter $k$ by minimizing the sum of out-of-sample prediction errors of these estimators:
$$
\widehat k^{VCV} = \arg\min_{k\in\mathcal K}\sum_{v=1}^V\sum_{i\in I_v} (Y_i - \widehat f_{k,v}(X_i))^2.
$$
The ultimate estimator of the function $f$ can be either the average estimator
$$
\widehat f_{AVCV}(\cdot) = \frac{1}{V}\sum_{v=1}^V \widehat f_{\widehat k^{VCV},v}(\cdot)
$$
or the full-sample estimator $\widehat f_{FVCV}(\cdot) = \widehat f_{\widehat k^{VCV}}(\cdot)$, with the estimators $\widehat f_k(\cdot)$ being described in \eqref{eq: series estimator}. The analysis of the average estimator can be found in \cite{LM12}, and the full-sample estimator can be analyzed using the arguments in \cite{CL21}, who studied the quantile version of the estimator $\widehat f_{FVCV}$.

The estimators $\widehat f_{AVCV}$ and $\widehat f_{FVCV}$ are more efficient then $\widehat f^V$ but are still not asymptotically optimal. To obtain an asymptotically optimal estimator, we should assume that the number of folds $V$ gets large together with the sample size $n$. In particular, the arguments in \cite{CL21} can be used to show that as $V$ gets large, the full sample estimator $\widehat f_{FVCV}$ becomes asymptotically optimal in the sense that it satisfies \eqref{eq: ideal result} in the $\ell_2$ metric. In the extreme case, when $V = n$, we obtain the leave-one-out cross-validation, which was studied in \cite{L87} in the context of the mean regression \eqref{eq: model} and \cite{C14} in the context of density estimation, among others. The problem with the leave-one-out cross-validation, however, is that it is computationally difficult, as it requires calculating $n$ estimators for each $k$. On the other hand, in the case of the series estimator, computation turns out to be not so difficult because of the existence of an explicit expression: when $V=n$, the expression for $\widehat k^{VCV}$ above reduces to
$$
\widehat k^{LOO} = \arg\min_{k\in\mathcal K}\frac{1}{n}\sum_{i=1}^n \frac{(Y_i - \widehat f_k(X_i))^2}{(1 - m_k)^2},
$$
where $m_k$ is the $k$th diagonal element of the matrix $\M_k$; see equation (1.5) in \cite{L87}.

More generally, cross-validation procedures can be applied in many settings, e.g. see \cite{BBL23}, but they always give results in the $\ell_2$ metric and can not be applied if we are interested in the uniform or pointwise metrics. See \cite{AC10} for a deep review of the theory underlying cross-validation procedures. See also \cite{L20} for more recent results.

\subsection{The Method of Penalization}\label{sec: penalization method}
The penalization method is based on the idea that the series estimator with too many terms $k$ suffers from overfitting the data, i.e. it yields a small value of the least squares criterion function but has large variance. To avoid choosing too many series terms, the penalization method adds a penalty to the least squares criterion function that is proportional to $k$:
\begin{equation}\label{eq: penalization method}
\widehat k^P = \arg\min_{k\in\mathcal K_n} \left\{ \frac{1}{n}\sum_{i=1}^n(Y_i - \widehat f_k(X_i))^2 +\frac{\lambda k}{n} \right\},
\end{equation}
where $\lambda>0$ is the penalty parameter.

The penalization method thus generalizes the Mallows method under homoskedasticity, which corresponds to setting $\lambda = 2\sigma^2$ in \eqref{eq: penalization method}; see the expression in \eqref{eq: mallows method homoskedasticity}. In fact, the Mallows method can be seen as a version of the penalization method where the penalty parameter $\lambda$ is chosen so that the resulting criterion function yields an unbiased risk estimator for the series estimator. 
The Mallows method yields the series estimator $\widehat f_{\widehat k^M}$ that is asymptotically optimal, and so setting $\lambda = 2\sigma^2$ under homoskedasticity indeed seems reasonable.
The benefit of having a flexible penalty parameter, however, becomes apparent when we generalize the problem of selecting the number of terms $k$ for the series estimator and consider the general problem of {\em model selection}. To describe it, consider a set of models $\mathcal M_n$ such that each model $m\in\mathcal M_n$ corresponds to some set of basis functions $\{p_{m,1},\dots,p_{m,k(m)}\}$. Then for each model $m\in\mathcal M_n$, we have a corresponding series estimator $\widehat f_m(\cdot)$ defined by analogy with \eqref{eq: series estimator} where we replace $p^k$ by $p^m = (p_{m,1},\dots,p_{m,k(m)})^{\top}$. We are then interested in constructing an estimator $\widehat m\in\mathcal M_n$ such that $d(\widehat f_m,f)$ is as small as possible in general and satisfies analogs of \eqref{eq: ideal result}, \eqref{eq: oracle probability}, or \eqref{eq: oracle expectation} in particular. The model selection problem generalizes the problem of selecting the number of terms $k$ for the series estimator in the sense that the latter corresponds to the model selection problem where for each $k\in\mathcal K_n$, we have only one model $m\in\mathcal M_n$ such that $k(m) = k$, and the models are assumed to be ordered in terms of their ability to approximate the true function $f$. The model selection problem also generalizes the {\em variable selection} problem, where we are given a set of basis functions (or variables) $\{p_1,\dots,p_{N_n}\}$ and each model $m$ corresponds to a subset of this set. We say that we have an {\em ordered variable selection} problem if the models are nested, e.g. $\mathcal M_n = \{1,\dots,N_n\}$ and each model $m$ corresponds to the basis functions $p_1,\dots,p_m$, and we say that we have a {\em complete variable selection} problem if all subsets of the set $\{p_1,\dots,p_{N_n}\}$ are included in the set of models. Clearly, the ordered variable selection problem is closely related to the problem of selecting the number of terms for the series estimator. In fact, two problems coincide if the basis functions $p_j$ in the vector $p^k$ are independent of $k$, which is the case when the basis functions are given by monomials, for example. 

It turns out that the penalization method can be extended to work with the model selection problem as well:
\begin{equation}\label{eq: penalized model selection estimator}
\widehat m = \arg\min_{m\in\mathcal M_n} \left\{\frac{1}{n}\sum_{i=1}^n(Y_i - \widehat f_m(X_i))^2 + \frac{\lambda_m |m|}{n}\right\},
\end{equation}
where $\lambda_m$'s are model specific penalty parameters and $|m|= k(m)$ denotes the number of basis functions in the model $m$. We will see below why it makes sense to consider model-specific penalty parameters. 

It is obvious that the Mallows, Stein, and cross-validation methods discussed above can all be extended in a straightforward manner to work with the model selection problem as well but the penalization approach is expected to work better if the set of models $\mathcal M_n$ is large. For example, the Mallows method for model selection under homoskedastisticity would set $\lambda_m = 2\sigma^2$ for all $m\in\mathcal M_n$ in \eqref{eq: penalized model selection estimator}, which makes the criterion function an unbiased risk estimator of the estimator $\widehat f_m$. However, as demonstrated in \cite{BM07}, this choice is inappropriate if we have too many models in the set $\mathcal M_n$. In particular, it is inappropriate if we are dealing with the complete variable selection problem. The issue here is that even though the criterion function yields an unbiased risk estimator for each model $m$, when we are dealing with many models simultaneously, it is likely to happen that at least for some models, the values of the criterion function are much smaller than the risk of the corresponding series estimators, yielding misleading risk estimates. In fact, when dealing with the complete variable selection problem, setting $\lambda_m = 2\sigma^2$ for all $m\in\mathcal M_n$ may be so bad that the resulting estimator $\widehat f_{\widehat m}$ may not be even consistent. Thus, the Mallows, Stein, and cross-validation methods, which are all based on the idea of unbiased risk estimation, may not be appropriate if the set $\mathcal M_n$ is too large in general and if we are dealing with the complete variable selection problem in particular.

To describe the properties of the penalized model selection estimator $\widehat m$ in \eqref{eq: penalized model selection estimator}, let $\{L_m\}_{m\in\mathcal M_n}$ be a sequence of non-negative numbers and let $\theta\in(0,1)$ and $\kappa>2 - \theta$ be some constants. Assuming Gaussian and homoskedastic noise, $e\mid X\sim N(0,\sigma^2)$, \cite{BM07} proved the following result: If we set
\begin{equation}\label{eq: penalty inequality}
\lambda_m \geq \sigma^2\Big(\kappa + 2(2-\theta)\sqrt{L_m} + 2\theta^{-1}L_m\Big),\quad\text{for all }m\in\mathcal M_n,
\end{equation}
then the estimator $\widetilde f = \widehat f_{\widehat m}$ satisfies
\begin{equation}\label{eq: birge massart bound}
 (1-\theta)\E\left[\|\widetilde f - f\|^2_{2,n}\right]  \leq \inf_{m\in\mathcal M_n}\bigg\{ \E\left[\|\widehat f_m - f\|^2_{2,n}\right] + \frac{(\lambda_m - 2\sigma^2)|m|}{n}\bigg\} + \varepsilon_n,
\end{equation}
where
$$
\varepsilon_n = \frac{\sigma^2}{n}\left(\frac{(2-\theta)^2}{\kappa + \theta - 2} + \frac{2}{\theta}\right)\sum_{m\in\mathcal M_n}e^{-L_m|m|}.
$$
Based on this extremely nuanced result, \cite{BM07} proposed the following choice of the penalty parameters:
\begin{equation}\label{eq: penalization penalty choice}
\lambda_m = 2\sigma^2\left(1 + 2\sqrt{\frac{\log H_m}{|m|}} + \frac{2\log H_m}{|m|}\right), \quad m\in\mathcal M_n,
\end{equation}
where $H_m$ is the number of models with $|m|$ basis functions.\footnote{See \cite{A19} for a review of methods to estimate $\sigma^2$ in the model selection framework, which can be used to make the choice \eqref{eq: penalization penalty choice} feasible.}

Indeed, setting $\theta = 1/2$, $\delta>0$ such that $3\sqrt \delta + 4\delta <1/2$, $\kappa>3/2$ such that $\kappa + 3\sqrt\delta + 4\delta < 2$, and $L_m = (\log H_m)/|m| + \delta$ for all $m\in\mathcal M_n$, we have that all inequalities in \eqref{eq: penalty inequality} with $\lambda_m$'s given by \eqref{eq: penalization penalty choice} are satisfied, and so it follows from \eqref{eq: birge massart bound} that
$$
\E\left[\|\widetilde f - f\|^2_{2,n}\right] \leq 2 \inf_{m\in\mathcal M_n} \left\{\E\left[\|\widehat f_m - f\|^2_{2,n}\right] + \frac{8\sigma^2|m|\log N_n}{n} \right\} + \frac{c\sigma^2}{n}
$$
for the complete variable selection and
\begin{equation}\label{eq: oracle inequality penalization ordered selection}
\E\left[\|\widetilde f - f\|^2_{2,n}\right] \leq 2 \inf_{m\in\mathcal M_n} \E\left[\|\widehat f_m - f\|^2_{2,n}\right]  + \frac{c\sigma^2}{n}
\end{equation}
for the ordered variable selection, where $c>0$ is some constant. These are oracle inequalities of the form \eqref{eq: oracle expectation} in the prediction norm.

Note also that the choice \eqref{eq: penalization penalty choice} yields the Mallows method for the ordered variable selection under homoskedasticity; compare \eqref{eq: penalization penalty choice} with $H_m = 1$ for all $m\in\mathcal M_n$ and \eqref{eq: mallows method homoskedasticity}. Moreover, the result \eqref{eq: birge massart bound} is sharp enough to recover the asymptotic optimality of the Mallows method in this case under the assumption that $\min_{m\in\mathcal M_n}n\E[\|\widehat f_m - f\|_{2,n}^2]\to \infty$. Indeed, setting $\theta = \theta_n$, $\kappa = \kappa_n = 2 - \theta_n/2$, $\delta = \delta_n$ such that $3\sqrt{\delta_n} + 4\delta_n = \theta_n / 4$, and $L_m = (\log H_m)/|m| + \delta_n$ for all $m\in\mathcal M_n$, it follows that as long as $\theta_n\to0$ sufficiently slowly, we have $\varepsilon_n / \min_{m\in\mathcal M_n}\E[\|\widehat f_m - f\|_{2,n}^2] \to 0$, yielding the asymptotic optimality result.



More generally, the penalization approach is popular in the empirical risk minimization framework, where the function $f$ is assumed to solve the optimization problem $f = \arg\min_{f\in\mathcal F}\E[m(Z,f)]$, and our task is to choose an estimator in the set $\{\widehat f_k = \arg\min_{f\in\mathcal F_k}n^{-1}\sum_{i=1}^n m(Z_i,f)\colon k\in\mathcal K_n\}$, where $\mathcal F_k$'s are subsets of $\mathcal F$. See \cite{M07} and \cite{K11} for detailed expositions of this framework.

\subsection{The Method of Aggregation}

Instead of selecting one estimator $\widehat f_{\widehat k}$ from the collection of estimators $\{\widehat f_k\colon k\in\mathcal K_n\}$, the aggregation method takes a weighted average of these estimators: $\widetilde f(\cdot) = \sum_{k\in\mathcal K_n} w_k \widehat f_k(\cdot)$ for some weights $w_k$. The idea here is that by taking a weighted average of the estimators, we can average out their estimation errors in the same fashion as diversifying portfolio helps to minimize its risk in finance. In this section, I will describe a weighting scheme due to \cite{LB06} that yields an estimator $\widetilde f$ with particularly attractive properties. Throughout this section, we will assume that the model is homoskedastic and Gaussian, i.e. $ e \mid X = N(0,\sigma^2)$.

To describe their weighting scheme, recall that as follows from our discussion in Section \ref{sub: mallows}, an unbiased estimator of the risk $\E[\|\widehat f_k - f\|_{2,n}^2]$ of the series estimator $\widehat f_k$ based on $k$ terms can be taken to be
$$
\widehat r_k = \frac{1}{n}\sum_{i=1}^n (Y_i - \widehat f_k(X_i))^2 + \frac{2\sigma^2 k}{n} - \sigma^2, \quad k\in\mathcal K_n.
$$
It is thus natural to give higher weights $w_k$ to estimators $\widehat f_k$ with smaller $\widehat r_k$. Indeed, \cite{LB06} proposed setting
\begin{equation}\label{eq: leung barron weights}
w_k^{LB} = \frac{\exp(-n\widehat r_k/(4\sigma^2))}{\sum_{k'\in\mathcal K_n}\exp(-n\widehat r_{k'}/(4\sigma^2))},\quad k\in\mathcal K_n.
\end{equation}
Note that the resulting estimator $\widetilde f^{LB}(\cdot) = \sum_{k\in\mathcal K_n}w^{LB}_k\widehat f_k(\cdot)$ is highly non-linear in $Y_i$'s but using the Stein method as described in Section \ref{sub: stein}, they showed that an unbiased estimator $\widetilde r^{LB}$ of the risk of $\widetilde f^{LB}$ can be taken to be
$\widetilde r^{LB} = \sum_{k\in\mathcal K_n} w_k^{LB}\widehat r_k,$
and this result can be used to derive an oracle inequality for $\widetilde f^{LB}$. Indeed, solving \eqref{eq: leung barron weights} for $\widehat r_k$, we have
\begin{equation}\label{eq: formula for risk from weights}
\widehat r_k = -\frac{4\sigma^2}{n}\left( \log w_k^{LB} + \log\left( \sum_{k'\in\mathcal K_n} \exp\left(-\frac{n\widehat r_{k'}}{4\sigma^2}\right) \right) \right).
\end{equation}
Also, denoting $k^* = \arg\min_{k\in\mathcal K_n} \widehat r_k$ and subtracting the expression \eqref{eq: formula for risk from weights} for $k = k^*$ from the same expression for general $k$,
$$
\widehat r_k - \widehat r_{k^*} = - \frac{4\sigma^2}{n}\left( \log w_k^{LB} - \log w_{k^*}^{LB} \right) \leq - \frac{4\sigma^2}{n}\log w_k^{LB}.
$$
Combining this inequality with $\widetilde r^{LB} = \sum_{k\in\mathcal K_n} w_k^{LB}\widehat r_k$, we obtain
$$
\widetilde r^{LB} \leq \widehat r_{k^*} - \frac{4\sigma^2}{n}\sum_{k\in\mathcal K_n} w_k^{LB}\log w_k^{LB} \leq \widehat r_{k^*} + \frac{4\sigma^2\log|\mathcal K_n|}{n}.
$$
Finally, taking expectations on both sides of this inequality, we obtain the main result from \cite{LB06}: if $e\mid X\sim N(0,\sigma^2)$, then
\begin{equation}\label{eq: leung barron bound}
\E\left[ \|\widetilde f^{LB} - f\|_{2,n}^2 \right]
\leq \min_{k\in\mathcal K_n} \E\left[  \|\widehat f_k - f\|_{2,n}^2 \right] + \frac{4\sigma^2\log |\mathcal K_n|}{n}.
\end{equation}
This result is remarkable because it yields an oracle inequality of the type \eqref{eq: oracle expectation} with the constant $C=1$. Compare, for example, this inequality with the oracle inequality \eqref{eq: oracle inequality penalization ordered selection} for the estimator $\widehat f_{\widehat k}$ based on the penalization method, where we had the constant $C=2$. In fact, the right-hand of \eqref{eq: leung barron bound} is typically smaller than the right-hand of \eqref{eq: oracle inequality penalization ordered selection}. In particular, as long as $|\mathcal K_n|\leq n$, the bound \eqref{eq: leung barron bound} is preferred as long as $k^* = \arg\min_{k\in\mathcal K_n} \E[\|\widehat f_k - f\|_{2,n}^2]$ is of larger order than $\log n$. Related results can be found, for example, in \cite{C97}, \cite{Y00, Y04}, and \cite{DT07}.




\subsection{Simulation Study}
In this section, I compare the performance of the methods described above via simulations. Since the Stein and penalization methods coincide in the case of the series estimator with the Mallows method, I omit them from consideration. I thus focus on the following four methods: Mallows, Lepski, cross-validation, and aggregation. For all methods, I consider the nonparametric mean regression model \eqref{eq: model} with the function $f$ being given by either $f(x) = \exp(\exp(x))$ or $f(x)=\sin(2\pi x)$ for all $x\in\R$. I assume that $X\sim U([0,1])$ and $e=\varepsilon/\sqrt{1+X^2}$, where $\varepsilon\mid X\sim N(0,1)$. Further, I consider random samples of size $n = 500$ and $n=1000$. I study the series estimator using either monomials or quadratic splines as described above. For both sets of basis functions, I set $\mathcal K_n = \{1,\dots,[n^{1/3}]\}$. For the Mallows, Lepski, and aggregation methods, I obtain regression residuals $\widehat e_i$ using $\widehat f_k$ with $k=[n^{1/3}]$. For the Lepski method, I set $x_0 = 0.5$. For the cross-validation method, I focus on the $V$-fold cross-validation with $V=5$. Finally, since the aggregation method as described in the previous section is aiming at estimating homoskesdatic model, I extend it in a somewhat ad-hoc way as follows. I set
$$
w_k = \frac{\exp(-n\widehat r_k/(4\widehat\sigma^2))}{\sum_{k'\in\mathcal K_n}\exp(-n\widehat r_{k'}/(4\widehat\sigma^2))},\quad\text{for all }k\in\mathcal K_n,
$$
where $\widehat \sigma^2 = n^{-1}\sum_{i=1}^n \widehat e_i^2$ and
$$
\widehat r_k = \frac{1}{n}\sum_{i=1}^n(Y_i - \widehat f_k(X_i))^2 + \frac{2\widehat A_k}{n} - \widehat\sigma^2,\quad\text{for all }k\in\mathcal K_n,
$$
is an unbiased estimator of the risk $\|\widehat f_k - f\|_{2,n}^2$ of the series estimator $\widehat f_k$ appearing in our discussion of the Mallows method; see in particular \eqref{eq: ak def} for the definition of $\widehat A_k$. All other specifications coincide with recommendations for individual methods above.

The simulation results are presented in Table 1. Overall, we observe that estimation results improve as we increase the sample size. Also, the aggregation method tends to give the best results. The Lepski method may fail to give good results in the $\ell_2$, prediction, and uniform metrics, see the case of monomials and $f(x) = \sin(2\pi x)$. This is not a bug but a feature of the method, as it is constructed specifically to yield good results in the pointwise metric, and it does perform very well in the pointwise metric. In fact, in the case of monomials and $f(x) = \sin(2\pi x)$, the Lepski method substantially outperforms the other methods in the pointwise metric. The Mallows and cross-validation methods have similar performance.

\newpage

\begin{center}
\begin{tabular}{ccccccccc}
\multicolumn{9}{c}{}\tabularnewline
\hline 
\hline 
\multicolumn{9}{c}{$n=500;\quad f(x)=\exp(\exp(x))$}\tabularnewline
\hline 
\hline 
method & $\ell_{2}$(M) & $\ell_{2,n}$(M) & $\ell_{\infty}$(M) & $\ell_{pw}$(M) & $\ell_{2}$(S) & $\ell_{2,n}$(S) & $\ell_{\infty}$(S) & $\ell_{pw}$(S)\tabularnewline
Mallows     &  0.100 &  0.096 &  0.303 &   0.071 &  0.105 &  0.101 &  0.316 &   0.078 \tabularnewline
Lepski      &  0.108 &  0.104 &  0.327 &   0.090 &  0.112 &  0.106 &  0.376 &   0.066 \tabularnewline
CV          &  0.098 &  0.094 &  0.291 &   0.069 &  0.104 &  0.099 &  0.322 &   0.072 \tabularnewline
Aggregation &  0.092 &  0.089 &  0.265 &   0.065 &  0.095 &  0.091 &  0.296 &   0.065 \tabularnewline
\hline 
\hline 
\multicolumn{9}{c}{$n=1000;\quad f(x)=\exp(\exp(x))$}\tabularnewline
\hline 
\hline 
method & $\ell_{2}$(M) & $\ell_{2,n}$(M) & $\ell_{\infty}$(M) & $\ell_{pw}$(M) & $\ell_{2}$(S) & $\ell_{2,n}$(S) & $\ell_{\infty}$(S) & $\ell_{pw}$(S)\tabularnewline
Mallows     &  0.071 &  0.069 &  0.223 &   0.045 &  0.078 &  0.076 &  0.227 &   0.052 \tabularnewline
Lepski      &  0.079 &  0.077 &  0.248 &   0.062 &  0.073 &  0.071 &  0.199 &   0.048 \tabularnewline
CV          &  0.072 &  0.070 &  0.227 &   0.048 &  0.077 &  0.074 &  0.220 &   0.049 \tabularnewline
Aggregation &  0.067 &  0.065 &  0.202 &   0.044 &  0.070 &  0.068 &  0.201 &   0.045 \tabularnewline
\hline 
\hline 
\multicolumn{9}{c}{$n=500;\quad f(x)=\sin(2\pi x)$}\tabularnewline
\hline 
\hline 
method & $\ell_{2}$(M) & $\ell_{2,n}$(M) & $\ell_{\infty}$(M) & $\ell_{pw}$(M) & $\ell_{2}$(S) & $\ell_{2,n}$(S) & $\ell_{\infty}$(S) & $\ell_{pw}$(S)\tabularnewline
Mallows     &  0.104 &  0.100 &  0.325 &   0.064 &  0.092 &  0.088 &  0.300 &   0.050 \tabularnewline
Lepski      &  0.448 &  0.438 &  0.962 &   0.029 &  0.091 &  0.087 &  0.250 &   0.048 \tabularnewline
CV          &  0.106 &  0.102 &  0.332 &   0.062 &  0.086 &  0.083 &  0.271 &   0.053 \tabularnewline
Aggregation &  0.094 &  0.091 &  0.276 &   0.058 &  0.090 &  0.086 &  0.300 &   0.048 \tabularnewline
\hline 
\hline 
\multicolumn{9}{c}{$n=1000;\quad f(x)=\sin(2\pi x)$}\tabularnewline
\hline 
\hline 
method & $\ell_{2}$(M) & $\ell_{2,n}$(M) & $\ell_{\infty}$(M) & $\ell_{pw}$(M) & $\ell_{2}$(S) & $\ell_{2,n}$(S) & $\ell_{\infty}$(S) & $\ell_{pw}$(S)\tabularnewline
Mallows     &  0.080 &  0.077 &  0.255 &   0.053 &  0.071 &  0.070 &  0.182 &   0.047 \tabularnewline
Lepski      &  0.434 &  0.424 &  0.948 &   0.031 &  0.076 &  0.074 &  0.204 &   0.054 \tabularnewline
CV          &  0.083 &  0.080 &  0.271 &   0.051 &  0.070 &  0.069 &  0.178 &   0.043 \tabularnewline
Aggregation &  0.073 &  0.071 &  0.217 &   0.049 &  0.067 &  0.066 &  0.180 &   0.043 \tabularnewline
\hline 
\hline 
\end{tabular}
\end{center}
\medskip
\noindent
Table 1: This table shows the average estimation error in the $\ell_2$ metric ($\ell_2$), prediction metric ($\ell_{2,n}$), uniform metric ($\ell_{\infty}$), and the pointwise metric ($\ell_{pw}$) of the series estimator based on either monomials (M) or quadratic splines (S) with the number of series terms determined by the Mallows, Lepski, CV, and Aggregation methods. CV means 5-fold cross-validation, and the three metrics are given by \eqref{eq: l2 metric}, \eqref{eq: linf metric}, and \eqref{eq: lpw metric} with the supremum in \eqref{eq: linf metric} being taken over $\mathcal X = [0,1]$ and $x_0$ in \eqref{eq: lpw metric} being equal to 0.5.

\newpage

\section{Tuning Parameter Selection for $\ell_1$-penalized Estimation}\label{sec: ell1 penalized estimation}

In this section, I discuss tuning parameter selection in the context of high-dimensional estimation via $\ell_1$-penalization. I discuss methods based on the theory of self-normalized moderate deviations, bootstrap, Stein's unbiased risk estimation, and cross-validation. I mainly focus on the case of i.i.d. data but describe extensions for the cases of clustered and panel data as well. Similarly, I mainly focus on the case of the mean regression but also touch the cases of the quantile regression and, more generally, generalized linear models.

Since the literature on high-dimensional estimation in general and on tuning parameter selection in high-dimensional settings in particular is large, my review here is rather selective. For example, I do not discuss methods such as the square-root Lasso, where the choice of the tuning parameter is relatively simple. An interested reader can find this and other existing methods, for example, in \cite{WW20}.

\subsection{Basic Lasso Theory}\label{sec: basic lasso theory}

Consider the following mean regression model:
\begin{equation}\label{eq: linear model}
Y = X^{\top}\beta + e,\quad \E[e\mid X] = 0,
\end{equation}
where $Y\in\R$ is a dependent variable, $X\in\mathbb R^p$ is a vector of covariates, $\beta\in\mathbb R^p$ is a vector of parameters, and $e\in\mathbb R$ is noise. Assume that we have a random sample $(X_1,Y_1),\dots,(X_n,Y_n)$ from the distribution of the pair $(X,Y)$. Assume also that the model is high-dimensional in the sense that $p$ is large, potentially larger or much larger than $n$, but sparse in the sense that the number of non-zero components of the vector $\beta$, say $s$, is small.

In the context of this model, it is well-known that the OLS estimator is rather poor and that the Lasso estimator is preferred:
\begin{equation}\label{eq: lasso definition}
\widehat \beta = \widehat\beta(\lambda) = \arg\min_{b\in\mathbb R^p} \left\{\frac{1}{n}\sum_{i=1}^n(Y_i - X_i^{\top}b)^2 + \lambda\|b\|_1\right\},
\end{equation}
where $\lambda> 0$ is a penalty parameter. In this section, I review the basic Lasso theory, as it guides the choices of $\lambda$ to be discussed below.\footnote{The discussion throughout Section \ref{sec: ell1 penalized estimation} assumes that the random variables in the vector $X$ are normalized so that their second moments are of the same order. If they are not normalized, it may be appropriate to consider the Lasso estimator of the form $\widehat\beta = \arg\min_{b\in\R^p}\{n^{-1}\sum_{i=1}^n (Y_i - X_i^{\top}b)^2 + \lambda \sum_{j=1}^p \kappa_j |b_j|\}$, where $\kappa_j$'s are appropriate loadings, e.g. see \cite{BCCH12}.}

Let $T = \{j=1,\dots,p\colon \beta_j\neq 0\}$ be the set of indices corresponding to non-zero components of the vector $\beta = (\beta_1,\dots,\beta_p)^{\top}$ and let $T^c = \{1,\dots,p\}\setminus T$ be the set of remaining indices. Also, for any vector $\delta = (\delta_1,\dots,\delta_p)^{\top}\in\mathbb R^p$, let $\delta_T = (\delta_{T1},\dots,\delta_{Tp})^{\top}\in\mathbb R^p$ be the vector defined by $\delta_{Tj} = \delta_j$ if $j\in T$ and $\delta_{Tj} = 0$ if $j\in T^c$ and let $\delta_{T^c}\in\mathbb R^p$ be the vector defined in the same way where the roles of $T$ and $T^c$ are switched. In addition, let 
$
\|\delta\|_{2,n} = \sqrt{n^{-1}\sum_{i=1}^n (X_i^{\top}\delta)^2}
$
denote the prediction norm of the vector $\delta\in\R^p$. Moreover, for any $c>1$, let $\mathcal R_c = \{\delta\in \R^p\colon \|\delta_{T^c}\|_1 \leq c\|\delta_T\|_1\}$ be the so-called {\em restricted set} and let
$$
\kappa_c = \inf_{\delta\in \mathcal R_c}\frac{\sqrt s\|\delta\|_{2,n}}{\|\delta_T\|_1}
$$
be the so-called {\em compatibility constant}. As we will see below, the compatibility constant plays an important role in the analysis of the Lasso estimator. Importantly, however, for any $c>1$, the compatibility constant $\kappa_{c}$ is bounded below from zero under so-called {\em sparse eigenvalue conditions}, which can be intuitively understood as requiring that there is no in-sample linear dependence between any $s\log n$ variables in the vector $X$; see \cite{BRT09} and \cite{BC11} for more precise descriptions.

Using the concept of the compatibility constants, we have the following result: For any $c>1$, if
\begin{equation}\label{eq: lambda constraint}
\lambda \geq 2c\max_{1\leq j\leq p}\left| \frac{1}{n}\sum_{i=1}^n X_{ij}\e_i \right|,
\end{equation}
then
\begin{equation}\label{eq: lasso bounds main}
\|\widehat \beta - \beta\|_{2,n} \leq \left(1 + \frac{1}{c}\right)\frac{\lambda\sqrt s}{\kappa_{\bar c}}\quad\text{and}\quad \|\widehat \beta - \beta\|_{1} \leq (1+\bar c)\left(1 + \frac{1}{c}\right)\frac{\lambda s}{\kappa_{\bar c}^2},
\end{equation}
where $\bar c = (c+1)/(c-1)$; see \cite{BRT09} and \cite{BC11}. A bound that is proportional to the bound for $\|\widehat\beta - \beta\|_{2,n}$ can also be derived, under certain conditions, for $\|\widehat \beta - \beta\|_2$; e.g. see \cite{BC10}. This result suggests the following general principle for choosing the penalty parameter $\lambda$: Choose $\lambda$ as small as possible subject to the constraint that \eqref{eq: lambda constraint} holds with probability close to one. For example, under the assumption of homoskedastic and Gaussian noise, i.e. if $e\mid X\sim N(0,\sigma^2)$ with some known $\sigma>0$, setting $\lambda$ to be equal to
\begin{equation}\label{eq: lambda brt}
\widehat\lambda^{BRT} = \frac{2c\sigma}{\sqrt n}\Phi^{-1}\left(1 - \frac{\alpha}{2p}\right)\max_{1\leq j\leq p}\sqrt{\frac{1}{n}\sum_{i=1}^n X_{ij}^2}
\end{equation}
for some $c>1$ and $\alpha\in(0,1)$, where $\Phi$ is the cdf of the $N(0,1)$ distribution, will ensure via the union bound that \eqref{eq: lambda constraint} holds with probability at least $1 - \alpha$. Thus, we can use the choice \eqref{eq: lambda brt} with some $\alpha$ close to zero and some $c>1$, which I refer to as the BRT method. In practice, it is recommended to set $\alpha = 0.1/\log(p\vee n)$ and $c = 1.1$, where $p\vee n$ denotes the maximum between $p$ and $n$; e.g. see \cite{BCCH12}.\footnote{The choice of the constant $c$ is actually rather tricky. Indeed, the formal result \eqref{eq: lambda constraint}-\eqref{eq: lasso bounds main} requires that the constant $c$ satisfies $c>1$. However, it is well-known via simulations that optimal performance of the Lasso estimator based on the penalty parameter $\lambda$ selected according to the BRT method \eqref{eq: lambda brt} is often achieved by values of $c$ satisfying $c<1$. This, on the one hand, motivates the choice $c=1.1$ but, on the other hand, also implies that it would be of great interest to better understand the behavior of the Lasso estimator when the constraint \eqref{eq: lambda constraint} is not satisfied.} In the next sections, I will describe methods that allow to choose $\lambda$ without imposing the assumption of homoskedastic and Gaussian noise.

To conclude this section, I note that under mild regularity conditions, the random variable on the right-hand side of \eqref{eq: lambda constraint} converges to zero with the rate $\sqrt{\log p /n}$. Thus, given that the compatibility constant $\kappa_{\bar c}$ is bounded away from zero under sparse eigenvalue conditions, it follows that, with a proper choice of the penalty parameter $\lambda$, under certain regularity conditions, the Lasso estimator satisfies
\begin{equation}\label{eq: lasso rate of convergence}
\|\widehat \beta - \beta\|_{2} = O_P\left(\sqrt{\frac{s\log p}{n}}\right)\quad\text{and}\quad \|\widehat \beta - \beta\|_{1} = O_P\left(\sqrt{\frac{s^2\log p}{n}}\right).
\end{equation}
Note also that a similar result can be obtained for the approximately sparse model, where the vector $\beta$ does not have to be sparse but can be well approximated by a sparse vector with $s$ non-zero components.

\subsection{Selection via Self-Normalization}
I now discuss how to choose the penalty parameter $\lambda$ for the Lasso estimator following the general principle described in the previous section but without assuming homoskedastic and Gaussian noise. In this section, I focus on the method based on the theory of self-normalized moderate deviations. In the next section, I will describe the method based on the bootstrap theory.

The method based on the theory of self-normalized moderate deviations was proposed in \cite{BCCH12}. This theory states that even if the random variables $X_{ij}e_i$ have only three finite moments, the ratio of the distribution function of the random variables
$$
T_j = \frac{\sum_{i=1}^n X_{ij} e_i}{\sqrt{\sum_{i=1}^n X_{ij}^2 e_i^2}}
$$
and the distribution function of $N(0,1)$ random variables converges to one not only pointwise, which would follow from the classic Central Limit Theorems, but also uniformly over sets $(-\bar x_n,\bar x_n)$ as long as $\bar x_n = o(n^{1/6})$:
\begin{equation}\label{eq: snmd bound}
\sup_{x\in(-\bar x_n,\bar x_n)} \left| \frac{\P(T_j > x)}{\P(N(0,1) > x)} - 1 \right|\to 0\quad\text{as }n\to\infty,
\end{equation}
e.g. see \cite{JSW03} for the original result or Lemma A.1 in \cite{BCCHKN18} for exposition.\footnote{The classic Berry-Esseen theorem leads to bounds on $|\P(T_j > x) - P(N(0,1)>x)|$. The result \eqref{eq: snmd bound} is much stronger, however, as it provides a bound on $|\P(T_j > x) - P(N(0,1)>x)|/P(N(0,1)>x)$ and $\P(N(0,1)>x)\to 0$ as $x\to\infty$.} Thus, by denoting
$$
\bar\lambda_n = \frac{2c}{\sqrt n}\Phi^{-1}\left(1 - \frac{\alpha}{2p}\right)\max_{1\leq j\leq p}\sqrt{\frac{1}{n}\sum_{i=1}^n X_{ij}^2 e_i^2},
$$
it follows via the union bound that
\begin{align}
& \P\left(2c\max_{1\leq j\leq p}\left| \frac{1}{n}\sum_{i=1}^n X_{ij}e_i \right| > \bar\lambda _n\right)
 \leq \sum_{j=1}^p \P\left( |T_j| > \Phi^{-1}\left( 1 - \frac{\alpha}{2p} \right) \right) \nonumber\\
& \qquad  \leq 2\sum_{j=1}^p \P\left(N(0,1) > \Phi^{-1}\left( 1 - \frac{\alpha}{2p} \right)\right) (1 + o(1))  = \alpha(1 + o(1))\label{eq: union bound derivation}
\end{align}
as long as $\log(p/\alpha) = o(n^{1/3})$ since $\Phi^{-1}(1 - \alpha/(2p)) \leq \sqrt{2\log(2p/\alpha)}$ by Proposition 2.5 in \cite{D14}. Note, however, that setting $\lambda = \bar\lambda_n$ is infeasible since we do not observe $e_i$'s. \cite{BCCH12} proposed the following three-step algorithm, which I refer to as the BCCH method, to work around this issue. First, we run Lasso with
\begin{equation}\label{eq: conservative lambda}
\lambda = \frac{2c}{\sqrt n}\Phi^{-1}\left(1 - \frac{\alpha}{2p}\right)\max_{1\leq j\leq p}\sqrt{\frac{1}{n}\sum_{i=1}^n X_{ij}^2 Y_i^2}
\end{equation}
to get a preliminary Lasso estimator $\widehat \beta^P$. Second, we calculate regression residuals $\widehat e_i = Y_i - X_i^{\top}\widehat\beta^P$ for all $i=1,\dots,n$. Third, we set $\lambda$ to be equal to
\begin{equation}\label{eq: non-conservative lambda}
\widehat\lambda^{BCCH} = \frac{2c}{\sqrt n}\Phi^{-1}\left(1 - \frac{\alpha}{2p}\right)\max_{1\leq j\leq p}\sqrt{\frac{1}{n}\sum_{i=1}^n X_{ij}^2 \widehat e_i^2}.
\end{equation}
The final Lasso estimator is then $\widehat\beta = \widehat\beta(\widehat\lambda^{BCCH})$. The idea here is that because $\sum_{i=1}^n \E[X_{ij}^2Y_i^2] > \sum_{i=1}^n \E[X_{ij}^2 e_i^2]$ for all $j=1,\dots,p$, which follows from \eqref{eq: linear model}, we have via the uniform law of large numbers that under certain regularity conditions,
\begin{equation}\label{eq: y-e variance}
\max_{1\leq j\leq p}\sqrt{\frac{1}{n}\sum_{i=1}^n X_{ij}^2 Y_i^2} > \max_{1\leq j\leq p}\sqrt{\frac{1}{n}\sum_{i=1}^n X_{ij}^2 e_i^2}
\end{equation}
with probability approaching one. Thus, the preliminary Lasso estimator $\widehat\beta^P$ satisfies the constraint \eqref{eq: lambda constraint} with probability at least $1 - \alpha + o(1)$ and, as long as $\alpha=\alpha_n \to 0$, yields consistent estimators $\widehat e_i$ of $e_i$'s. On the other hand, the preliminary Lasso estimator is conservative: because of \eqref{eq: y-e variance}, its penalty parameter $\lambda$ is larger than necessary. This issue is fixed on the third step, where we replace $Y_i$'s by $\widehat e_i$'s in the penalty choice $\lambda$; compare \eqref{eq: conservative lambda} and \eqref{eq: non-conservative lambda}.

Similarly to our discussion of the BRT method, for both the preliminary and the final Lasso estimators, \cite{BCCH12} suggested setting $\alpha = 0.1/\log(p\vee n)$ and $c = 1.1$. They proved that under certain regularity conditions, including the sparse eigenvalue conditions and the growth condition $\log p = o(n^{1/3})$, which appeared above because of the requirements of the theory of self-normalized moderate deviations, the final Lasso estimator $\widehat \beta = \widehat\beta(\widehat\lambda^{BCCH})$ satisfies \eqref{eq: lasso rate of convergence}. \cite{KPS24} extended this result to the case of a high-dimensional VAR model.

\subsection{Selection via Bootstrap}\label{sec: bootstrap lasso}
The BCCH method described in the previous section is still somewhat conservative: the third step of the BCCH method eliminates the slackness arising from \eqref{eq: y-e variance} but there is still slackness arising from our use of the union bound in \eqref{eq: union bound derivation}. Indeed, as long as the variables in the vector $X$ are correlated, a value of the penalty parameter $\lambda$ that is smaller than that produced by the BCCH method may be sufficient to satisfy constraint \eqref{eq: lambda constraint} with probability close to one. \cite{BC10} illustrated this point via simulations. In turn, smaller values of $\lambda$ will lead to tighter bounds on the estimation error in \eqref{eq: lasso bounds main} and are thus preferable. In this section, I describe a bootstrap method that gives $\lambda$ such that the probability of \eqref{eq: lambda constraint} is {\em equal} to $1-\alpha + o(1)$. 

The method consists of four steps. First and second steps are exactly the same as in the previous section: we first run Lasso with $\lambda$ given in \eqref{eq: conservative lambda} to obtain the preliminary Lasso estimator $\widehat \beta^P$ and then calculate regression residuals $\widehat e_i = Y_i - X_i^{\top}\widehat\beta^P$ for all $i=1,\dots,n$. On the third step, we estimate the $1-\alpha$ quantile of the random variable on the right-hand side of \eqref{eq: lambda constraint} via multiplier bootstrap. Specifically, let $w_1,\dots,w_n$ be i.i.d. $N(0,1)$ random variables that are independent of the data and let $\widehat q_{1-\alpha}$ be the $1-\alpha$ quantile of the conditional distribution of
\begin{equation}\label{eq: bootstrap stat max norm}
\max_{1\leq j\leq p}\left| \frac{1}{\sqrt n}\sum_{i=1}^n w_i X_{ij} \widehat e_i \right|
\end{equation}
given the data $\{(X_i,Y_i)\}_{i=1}^n$, and thus given $\{(X_i,\widehat e_i)_{i=1}^n\}$. The estimator of the $1-\alpha$ quantile of the random variable on the right-hand side of \eqref{eq: lambda constraint} is then $\widehat q_{1-\alpha}/\sqrt n$. On the fourth step, we set $\lambda$ to be equal to $\widehat\lambda^{B} = (2c/\sqrt n)\widehat q_{1-\alpha}$, so that the final Lasso estimator is $\widehat \beta = \widehat\beta(\widehat\lambda^B)$, where we again choose $\alpha = 0.1/\log(p\vee n)$ and $c=1.1$. 

The idea behind the bootstrap method is that conditional on $\{(X_i,\widehat e_i)_{i=1}^n\}$, the distribution of the vector $n^{-1/2}\sum_{i=1}^n w_i X_i \widehat e_i$ is centered Gaussian with covariance matrix $n^{-1}\sum_{i=1}^n \widehat e_i^2 X_iX_i^{\top}$, and so, using the Gaussian comparison inequalities, as developed in \cite{CCK15}, it is possible to show that the distribution of the max norm \eqref{eq: bootstrap stat max norm} of this vector can be well approximated by the distribution of the max norm of the centered Gaussian vector with covariance matrix $\E[e^2XX^{\top}]$. In turn, it follows from the high-dimensional Central Limit Theorems as developed in \cite{CCK13, CCK17, CCKK22} that the latter distribution also provides a good approximation to the distribution of the max norm of the vector $n^{-1/2}\sum_{i=1}^n X_{i}e_i$, e.g. see \cite{BCCHKN18} for an exposition of these results. Formally, one can show that using $\lambda = \widehat\lambda^B = (2c/\sqrt n)\widehat q_{1-\alpha}$ ensures that \eqref{eq: lambda constraint} holds with probability $1 - \alpha + o(1)$ as long as the growth condition $\log p = o(n^{1/5})$ and other regularity conditions are satisfied, which in turn implies that the Lasso estimator $\widehat\beta = \widehat\beta(\widehat\lambda^B)$ satisfies \eqref{eq: lasso rate of convergence}.

\subsection{Selection via Stein Method}\label{sec: Stein lasso}
Another approach for choosing the penalty parameter $\lambda$ for the Lasso estimator \eqref{eq: lasso definition} is based on the Stein method, as developed in \cite{ZHT07} and \cite{TT12}. The method takes a particularly simple form if we assume that the distribution of the vector $X$ is absolutely continuous with respect to the Lebesgue measure on $\R^p$, as the Lasso estimator is almost surely unique in this case. Assuming also that $e\mid X\sim N(0,\sigma^2)$, their method consists of setting $\lambda$ to be equal to
\begin{equation}\label{eq: lasso stein}
\widehat\lambda^S = \arg\min_{\lambda>0} \left\{ \frac{1}{n}\sum_{i=1}^n (Y_i - X_i^{\top}\widehat\beta(\lambda))^2 + \frac{2\sigma^2\|\widehat\beta(\lambda)\|_0}{n} - \sigma^2 \right\},
\end{equation}
where we use $\|\widehat\beta(\lambda)\|_0$ to denote the number of non-zero components of the vector $\widehat\beta(\lambda)$. As proven in \cite{ZHT07} for the low-dimensional case and in \cite{TT12} for the high-dimensional case, the criterion function in \eqref{eq: lasso stein} is an unbiased estimator of the risk of the Lasso estimator $\|\widehat\beta(\lambda) - \beta\|_{2,n}^2$, where we condition on $X_1,\dots,X_n$.\footnote{This method can also be referred to as the selection via degrees of freedom, as $\|\widehat\beta(\widetilde\lambda)\|_0$ turns out to be an unbiased estimator of the degrees of freedom of the Lasso estimator $\widehat\beta(\widehat\lambda)$; see \cite{ZHT07} and \cite{TT12}. Also, as proven in \cite{ZHT07}, the minimum in \eqref{eq: lasso stein} is achieved by one of the threshold values of $\lambda$, which are values such that as we increase $\lambda$, crossing them will change the set of non-zero components of the estimator $\widehat\beta(\lambda)$. The latter observation makes the problem of solving \eqref{eq: lasso stein} feasible.}

Intuitively, the KKT conditions for the optimization problem in \eqref{eq: lasso definition} are
$$
\frac{2}{n}\sum_{i=1}^n X_{ij}(Y_i - X_i^{\top}\widehat\beta(\lambda)) = \lambda\gamma_j,\quad\text{for all }j=1,\dots,p,
$$
where $\gamma = (\gamma_1,\dots,\gamma_p)^{\top}$ is a vector such that for all $j=1,\dots,p$, we have $\gamma_j = 1$ if $\widehat\beta_j(\lambda) > 0$, $\gamma_j = -1$ if $\widehat\beta_j(\lambda)<0$, and $\gamma_j\in[-1,1]$ if $\widehat\beta_j(\lambda)=0$.
Therefore, denoting $\Y = (Y_1,\dots,Y_n)^{\top}$ and $\widehat T = \{j=1,\dots,p\colon \widehat\beta_j(\lambda)\neq 0\}$  and letting $\X_{\widehat T}$ be the matrix consisting of the columns of $\X = (X_1,\dots,X_n)^{\top}$ corresponding to indices in $\widehat T$ and $\gamma_{\widehat T}$ be the vector consisting of the elements of $\gamma$ corresponding to indices in $\widehat T$, it follows that the Lasso estimator of the vector $\f = (X_1^{\top}\beta,\dots,X_n^{\top}\beta)^{\top}$ is
$$
\widehat\f = \X\widehat\beta(\lambda) = \X_{\widehat T}(\X_{\widehat T^{\top}}\X_{\widehat T})^{-1}(\X_{\widehat T}\Y - n\lambda \gamma_{\widehat T}/2).
$$
Hence, expression \eqref{eq: lasso stein} appears from expression \eqref{eq: stein definition} by taking the derivatives of $\widehat\f$ with respect to $\Y$ as discussed in Section \ref{sub: stein} and noting that both $\widehat T$ and $\gamma_{\widehat T}$ are locally independent of $\Y$ almost surely (in the sense that for almost all $\Y$, small perturbations of $\Y$ do not lead to any changes in $\widehat T$ and $\gamma_{\widehat T}$). 

The Lasso estimator $\widehat\beta = \widehat\beta(\widehat\lambda^S)$ is potentially attractive but its main disadvantage is that the choice \eqref{eq: lasso stein} is only justified under the assumption of homoskedastic and Gaussian noise. In addition, relative to the Mallows method of choosing the number of series terms, the analysis here requires studying concentration properties of the random variables $\|\widehat\beta(\lambda)\|_0$, which is difficult; compare \eqref{eq: mallows method homoskedasticity} and \eqref{eq: lasso stein}. \cite{BZ21} derived a bound for the variance of $\|\widehat\beta(\lambda)\|_0$ but, to the best of our knowledge, it is still not clear if the Lasso estimator $\widehat\beta = \widehat\beta(\widehat\lambda^S)$ satisfies \eqref{eq: lasso rate of convergence}.

\subsection{Selection via Cross-Validation}\label{sec: lasso with cross-validation}
The penalty parameter $\lambda$ for the Lasso estimator \eqref{eq: lasso definition} can also be selected by $V$-fold cross-validation.\footnote{In principle, I could consider here the leave-one-out cross-validation as well. However, even though the Lasso estimator is rather fast to compute, the leave-one-out cross-validation requires computing the Lasso estimator $n$ times, which does typically take too much time.} The algorithm proceeds as follows. Let $\Lambda_n$ be a set of candidate values of $\lambda$. This set is assumed to be large enough to contain a good choice of $\lambda$. In practice, this set can be chosen to consist of a geometric progression starting from some small value, say $1/n$, and ending at some large value, say $n$. Further, let $(I_1,\dots,I_V)$ be a random partition of the whole sample $I_0 = \{1,\dots,n\}$ into $V$ subsamples of roughly the same size. For each $\lambda \in \Lambda_n$ and $v=1,\dots,V$, we build the Lasso estimator of the vector $\beta$ using all the data excluding observations in subsample $v$:
$$
\widehat\beta(\lambda,v) = \arg\min_{b\in\R^p}\left\{\frac{1}{|I_0\setminus I_v|}\sum_{i\in I_0\setminus I_v}(Y_i - X_i^{\top}b)^2 + \lambda\|b\|_1\right\}.
$$
We then choose the penalty parameter $\lambda$ to be equal to a minimizer of the sum of out-of-sample prediction errors of these estimators:
\begin{equation}\label{eq: vcv lasso}
\widehat \lambda^{VCV} = \arg\min_{\lambda\in\Lambda_n}\sum_{v=1}^V\sum_{i\in I_v} (Y_i - X_i^{\top}\widehat\beta(\lambda,v))^2.
\end{equation}
The final Lasso estimator can be either the average estimator
$$
\widehat \beta^{AVCV} = \frac{1}{V}\sum_{v=1}^V \widehat\beta(\widehat\lambda^{VCV},v)
$$
or the full-sample estimator $\widehat\beta^{FVCV} = \widehat \beta(\widehat\lambda^{VCV})$, with the estimators $\widehat\beta(\lambda)$ being defined in \eqref{eq: lasso definition}.

\cite{LM12} derived an oracle inequality for the average estimator $\widehat\beta^{AVCV}$. Intuitively, the definition \eqref{eq: vcv lasso} implies that for any $\lambda\in\Lambda_n$,
\begin{equation}\label{eq: cv basic inequality lasso}
\sum_{v=1}^V\sum_{i\in I_v} (Y_i - X_i^{\top}\widehat\beta(\widehat\lambda^{VCV},v))^2 \leq \sum_{v=1}^V\sum_{i\in I_v} (Y_i - X_i^{\top}\widehat\beta(\lambda,v))^2
\end{equation}
Also, by the uniform law of large numbers and independence of $\{(X_i,Y_i)\}_{i\in I_v}$ from $\widehat\beta(\lambda,v)$, under certain regularity conditions,
\begin{equation}\label{eq: lln lasso cv}
\frac{1}{|I_v|}\sum_{i\in I_v}(Y_i - X_i^{\top}\widehat\beta(\lambda,v))^2 = \E_{X,Y}[(Y - X^{\top}\widehat\beta(\lambda,v))^2] + O_P\left(\sqrt{\frac{\log|\Lambda_n|}{n}}\right),
\end{equation}
uniformly over $\lambda\in\Lambda_n$, where $\E_{X,Y}[\cdot]$ denotes expectation with respect to the pair $(X,Y)$ holding the data fixed. Thus, combining \eqref{eq: cv basic inequality lasso} and \eqref{eq: lln lasso cv}, noting that $\E_{X,Y}[(Y - X^{\top}\widehat\beta(\lambda,v))^2] = \E[e^2] + E_{X,Y}[(X^{\top}(\beta - \widehat\beta(\lambda,v)))^2]$, and assuming for simplicity that $|I_1|=\dots=|I_V|$, we obtain
$$
\sum_{v=1}^V \E_{X,Y}[(X^{\top}(\widehat\beta(\widehat\lambda^{VCV},v) - \beta)^2] \leq \sum_{v=1}^V \E_{X,Y}[(X^{\top}(\widehat\beta(\lambda,v) - \beta)^2] + O_P\left(\sqrt{\frac{\log|\Lambda_n|}{n}}\right).
$$
uniformly over $\lambda\in\Lambda_n$. Here, all $V$ terms under the sum on the right-hand side are the same under our i.i.d. data assumption. Thus, by the triangle and Cauchy-Schwarz inequalities,
\begin{align*}
\E_{X,Y}[(X^{\top}(\widehat\beta^{AVCV} - \beta))^2]
& \leq \frac{1}{V}\sum_{v=1}^V \E_{X,Y}[(X^{\top}(\widehat\beta(\widehat\lambda^{VCV},v) - \beta)^2] \\
& \leq \min_{\lambda\in\Lambda_n}\E_{X,Y}[(X^{\top}(\widehat\beta(\lambda,1) - \beta)^2] + O_P\left(\sqrt{\frac{\log|\Lambda_n|}{n}}\right).
\end{align*}
Moreover, using a more careful argument, \cite{LM12} were able to get rid of the square root in the $O_P(\cdot)$ term in expense for having some extra terms in the inequality. Their oracle inequality implies in particular that, under certain regularity conditions, the average estimator $\widehat\beta = \widehat\beta^{AVCV}$ satisfies the $\ell_2$ norm bound in \eqref{eq: lasso rate of convergence}.

The analysis of the full-sample estimator $\widehat\beta^{FVCV}$ is more complicated. The key issue here is that the basic inequality \eqref{eq: cv basic inequality lasso} does not provide any direct information regarding this estimator. Moreover, simulation evidence suggests that the penalty parameter value selected by cross-validation, $\lambda = \widehat\lambda^{VCV}$, typically does not satisfy the constraint \eqref{eq: lambda constraint}, making the basic Lasso theory described in Section \ref{sec: basic lasso theory} inapplicable. To deal with these issues, \cite{CLC21} derived a relationship between the estimation error of the full-sample Lasso estimators $\widehat\beta(\lambda)$ and the corresponding subsample Lasso estimators $\widehat\beta(\lambda,v)$. The relationship turns out to depend on the estimation errors of the subsample Lasso estimators in the $\ell_1$ norm $\|\widehat\beta(\lambda,v) - \beta\|_1$, which can not be bounded using the Lecue and Mitchell arguments directly. Instead, \cite{CLC21} derived bounds on $\|\widehat\beta(\lambda,v) - \beta\|_1$ using the degrees of freedom formula for the Lasso estimator derived in \cite{ZHT07} and \cite{TT12}. As a result, they proved that, under certain regularity conditions, the full-sample estimator $\widehat\beta = \widehat\beta^{FVCV}$ satisfies
\begin{equation}\label{eq: clc bounds}
\|\widehat \beta - \beta\|_{2} = O_P\left(\sqrt{\frac{s(\log p)^2}{n}}\right)\quad\text{and}\quad \|\widehat \beta - \beta\|_{1} = O_P\left(\sqrt{\frac{s^2(\log p)^5}{n}}\right).
\end{equation}
These bounds, however, are by some $\log p$ factors worse than those in \eqref{eq: lasso rate of convergence}, which are guaranteed for the Lasso estimator with the penalty parameter selected either by the BCCH method or by the bootstrap method. It is therefore not clear whether the bounds \eqref{eq: clc bounds} are sharp.

\subsection{Selection under Clustering Dependence}\label{sec: clustered data}
In this section, I point out that one should be careful with methods described above if the data is clustered. To make this point clearly, suppose that $n$ observations $\{1,\dots,n\}$ can be split into $G$ groups such that there are $L$ observations within each group, so that $n=GL$, and all observations within the same group are perfectly correlated, so that $X_{ij}e_i = X_{lj}e_l$ whenever observations $i$ and $l$ belong to the same group. Then, letting $I(g)$ be a function that selects one observation from each group $g=1,\dots,G$, it follows that the constraint \eqref{eq: lambda constraint} reduces to
$$
\lambda \geq 2c\max_{1\leq j\leq p}\left| \frac{1}{G}\sum_{g=1}^G X_{I(g) j}e_{I(g)} \right|.
$$
Hence, for example, the BRT rule, which applies when $e\mid X\sim N(0,\sigma^2)$ with known $\sigma$, yields
\begin{align}
\lambda 
& = \frac{2c\sigma}{\sqrt G}\Phi^{-1}\left(1 - \frac{\alpha}{2p}\right)\max_{1\leq j\leq p}\sqrt{\frac{1}{G}\sum_{g=1}^G X_{I(g) j}^2} \nonumber \\
& = \frac{2c\sigma\sqrt L}{\sqrt n}\Phi^{-1}\left(1 - \frac{\alpha}{2p}\right)\max_{1\leq j\leq p}\sqrt{\frac{1}{n}\sum_{i=1}^n X_{i j}^2}. \label{eq: correct expression under clustering}
\end{align}
This expression should be compared with that in \eqref{eq: lambda brt}, which would be used by the researcher who ignores the clustering in the data. In particular, the correct expression \eqref{eq: correct expression under clustering} is $\sqrt L$ times larger, and using the wrong expression \eqref{eq: lambda brt} would yield the value of the penalty parameter $\lambda$ that does not necessarily satisfies \eqref{eq: lambda constraint} with probability at least $1-\alpha$. As a result, the researcher ignoring clustering could obtain misleading estimates that are far away from the true vector $\beta$.

Fortunately, it is possible to fix this issue as long as the number of clusters/groups is large. For concreteness, I focus here on the BCCH method, as studied in \cite{C18}. To adjust this method so that it works with clustered data, suppose as above that $n$ observations $\{1,\dots,n\}$ can be split into $G$ groups, that observations within each group can be arbitrarily correlated, and that observations across groups are independent. Letting $N(g)$ denote all observations $i$ within each group $g=1,\dots,G$, we then consider the following cluster-robust version of the BCCH method. First, we run Lasso with 
$$
\lambda = \frac{2c}{\sqrt n}\Phi^{-1}\left(1 - \frac{\alpha}{2p}\right)\max_{1\leq j\leq p}\sqrt{\frac{1}{n}\sum_{g=1}^G\left(\sum_{i\in N(g)} X_{ij} Y_i\right)^2}
$$
to get a preliminary Lasso estimator $\widehat \beta^P$. Second, we calculate regression residuals $\widehat e_i = Y_i - X_i^{\top}\widehat\beta^P$ for all $i=1,\dots,n$. Third, we set $\lambda$ to be equal to
$$
\widehat\lambda^{BCCH,C} = \frac{2c}{\sqrt n}\Phi^{-1}\left(1 - \frac{\alpha}{2p}\right)\max_{1\leq j\leq p}\sqrt{\frac{1}{n}\sum_{g=1}^G\left(\sum_{i\in N(g)} X_{ij} \widehat e_i\right)^2}
$$
The final Lasso estimator is then $\widehat\beta = \widehat\beta(\widehat\lambda^{BCCH,C})$. Using the same ideas as those in \cite{BCCH12}, it is then straightforward to show that this estimator satisfies \eqref{eq: lasso rate of convergence} with $n$ replaced by $n/L$, where $L = \max_{1\leq g\leq G}|N(g)|$ denotes the maximal number of observations within groups.

\subsection{Selection for Panel Data}

In this section, I consider the Lasso estimator in the context of panel data, where the dependent variable $Y_{it}\in \R$ for unit $i$ at time period $t$ satisfies the model
$$
Y_{it} = X_{it}^{\top}\beta + \xi_i + e_{it},\quad i=1,\dots,N, \ t=1,\dots,T,
$$
where $X_{it}\in\R^p$ is a vector of covariates, $\beta\in\mathbb R^p$ is a vector of parameters, $\alpha_i\in\R$ is an individual fixed effect that can be correlated with covariates $X_{it}$'s, and $e_{it}\in\R$ is noise that satisfies $\E[e_{it}\mid X_{i1},\dots,X_{iT}] = 0$. I assume that observations are independent across units and that the total number of observable periods $T$ is small.

The Lasso version of the classic fixed-effect estimator in this model takes the following form:
\begin{equation}\label{eq: lasso panel}
\widehat \beta = \arg\min_{b\in\R^p} \left\{\frac{1}{NT}\sum_{i=1}^N\sum_{t=1}^T (\widetilde Y_{it} - \widetilde X_{it}^{\top}b)^2 + \lambda\|b\|_1\right\},
\end{equation}
where $\widetilde Y_{it} = Y_{it} - T^{-1}\sum_{s=1}^T Y_{is}$ and $\widetilde X_{it} = X_{it} - T^{-1}\sum_{s=1}^T X_{is}$ for all $i=1,\dots,N$ and $t=1,\dots,T$. Using our discussion in Section \ref{sec: basic lasso theory}, it is then clear that a good choice of the penalty parameter $\lambda$ in \eqref{eq: lasso panel} should yield
\begin{equation}\label{eq: lambda constraint panel}
\lambda \geq 2c\max_{1\leq j\leq p}\left| \frac{1}{NT}\sum_{i=1}^N\sum_{t=1}^T \widetilde X_{itj} \widetilde e_{it} \right|,
\end{equation}
with probability close to one, where $\widetilde e_{it} = e_{it} - T^{-1}\sum_{s=1}^T e_{is}$ for all $i=1,\dots,N$ and $t=1,\dots,T$; compare \eqref{eq: lambda constraint panel} with \eqref{eq: lambda constraint}. In turn, using independence of observations across units and our discussion of clustered data in the previous section, it is straightforward to show that the panel version, for example, of the BCCH method takes the following form. First, we run Lasso with
$$
\lambda = \frac{2c}{\sqrt{NT}}\Phi^{-1}\left(1 - \frac{\alpha}{2p}\right)\max_{1\leq j\leq p}\sqrt{\frac{1}{NT}\sum_{i=1}^N\left(\sum_{t=1}^T \widetilde X_{itj} \widetilde Y_{it}\right)^2}
$$
to get a preliminary Lasso estimator $\widehat \beta^P$. Second, we calculate regression residuals $\widehat e_{it} = \widetilde Y_{it} - \widetilde X_{it}^{\top}\widehat\beta^P$ for all $i=1,\dots,N$ and $t=1,\dots,T$. Third, we set $\lambda$ to be equal to
$$
\widehat\lambda^{BCCH,P} = \frac{2c}{\sqrt{NT}}\Phi^{-1}\left(1 - \frac{\alpha}{2p}\right)\max_{1\leq j\leq p}\sqrt{\frac{1}{NT}\sum_{i=1}^N\left(\sum_{t=1}^T \widetilde X_{itj} \widehat e_{it}\right)^2}.
$$
The final Lasso estimator is then $\widehat\beta = \widehat\beta(\widehat\lambda^{BCCH,P})$. Using the same arguments as those in \cite{BCCH12}, one can show that this estimator satisfies \eqref{eq: lasso rate of convergence} with $n$ replaced by $N$.



\subsection{Pivotal Selection in Quantile Model}
In this section, I briefly discuss estimation of the high-dimensional quantile regression model
$$
Y = X^{\top}\beta + e,\quad Q_u(e\mid X) = 0,
$$
where $Q_u(e\mid X)$ is the $u$th quantile of the conditional distribution of $e$ given $X$, which is assumed to be continuous, and $u\in(0,1)$ is a fixed quantile index. As before, I assume that $\beta\in\R^p$ is high-dimensional but sparse: the number of its non-zero components is equal to some small $s$. If $(X_1,Y_1),\dots,(X_n,Y_n)$ is a random sample from the distribution of the pair $(X,Y)$, the analog of Lasso for this quantile model is the following $\ell_1$-penalized quantile regression estimator:
$$
\widehat \beta = \arg\min_{b\in\R^p} \left\{\frac{1}{n}\sum_{i=1}^n \rho_u(Y_i - X_i^{\top}b) + \lambda\|b\|_1\right\},
$$
where $t\mapsto \rho_u(t) = (u - 1\{t<0\})t$ is Koenker's check function and $1\{\cdot\}$ is the indicator function. \cite{BC11} proved the following result: if
\begin{equation}\label{eq: lambda quantile constraint}
\lambda \geq c\max_{1\leq j\leq p}\left| \frac{1}{n}\sum_{i=1}^n \frac{X_{ij}(u - 1\{u_i\leq u\})}{\sqrt{u(1-u)}} \right|,
\end{equation}
then
\begin{equation}\label{eq: quantile lasso bounds main}
\|\widehat \beta - \beta\|_2 \leq C\lambda\sqrt s\quad\text{and}\quad\|\widehat\beta - \beta\|_1 \leq C\lambda s,
\end{equation}
where $C>0$ is a constant that is independent of $(n, p, s, \lambda)$, $u_i = F(e_i\mid X_i)$ for all $i=1,\dots,n$, and $F(t\mid X) = \P(e\leq t\mid X)$ is the cdf of the conditional distribution function of $e$ given $X$. This result in analogous to that in \eqref{eq: lambda constraint}-\eqref{eq: lasso bounds main}, so that the same general principle for selecting the penalty parameter $\lambda$ applies, but the key point here is that $u_i\mid X_i$ has the $U(0,1)$ distribution, and so the conditional distribution of the random variable on the right-hand side of \eqref{eq: lambda quantile constraint} given $X_i$'s is pivotal and can be simulated exactly. We can thus choose the penalty parameter $\lambda$ as $cq_{1-\alpha}/\sqrt n$ with $q_{1-\alpha}$ being the $1-\alpha$ quantile of the $\sqrt n$ times the random variable on the right-hand side of \eqref{eq: lambda quantile constraint}, where $\alpha = 0.1/\log(p\vee n)$ and $c=1.1$ as before.

\subsection{Bootstrap-after-cross-validation in Generalized Linear Models}

Finally, I consider the generalized linear model
\begin{equation}\label{eq: generalized linear model}
\beta = \arg\min_{b\in\mathbb R^p}\E[m(X^{\top}b,Y)],
\end{equation}
where $m\colon \R\times\mathcal Y\to \R$ is a known function that is convex in its first argument and $\mathcal Y$ denotes the support of $Y$. Although both mean and quantile regression models fit into \eqref{eq: generalized linear model}, I am particularly interested here in other econometric models fitting into \eqref{eq: generalized linear model}, e.g. the logit model
$$
\beta = \arg\min_{b\in\R^p}\E[\log(1+\exp(X^{\top}b)) - YX^{\top} b]
$$
and the probit model
$$
\beta = \arg\min_{b\in\R^p}\E[-Y\log\Phi(X^{\top}b) - (1-Y)\log(1-\Phi(X^{\top}b))].
$$
As before, assuming that $\beta$ is high-dimensional but sparse and that we have a random sample $(X_1,Y_1),\dots,(X_n,Y_n)$ from the distribution of the pair $(X,Y)$, the analogue of Lasso for this framework is the following $\ell_1$-penalized M-estimator:
\begin{equation}\label{eq: penalized m-estimator}
\widehat\beta = \arg\min_{b\in\R^p}\left\{ \frac{1}{n}\sum_{i=1}^n m(X_i^{\top}b,Y_i) + \lambda\|b\|_1 \right\}.
\end{equation}
Applying the same ideas as those in the Lasso theory, we can show that if
\begin{equation}\label{eq: lambda m-estimation constraint}
\lambda \geq c\max_{1\leq j\leq p}\left|\frac{1}{n}\sum_{i=1}^n m_1(X_i^{\top}\beta,Y_i)X_{ij}\right|,
\end{equation}
then the result of the form \eqref{eq: quantile lasso bounds main} holds. We thus again want to choose the penalty parameter $\lambda$ as small as possible subject to the constraint that \eqref{eq: lambda m-estimation constraint} is satisfied with probability close to one. The main issue here, however, is that there may not exist a preliminary estimator allowing us to consistently estimate $\beta$ in the first step, as we did in the BCCH and bootstrap methods for the Lasso estimator. \cite{CS21} circumvented this problem and proposed obtaining the preliminary estimator using the subsample versions of the $\ell_1$-penalized estimator \eqref{eq: penalized m-estimator} with the penalty parameter $\lambda$ selected by cross-validation, analogous to estimators $\widehat\beta(\widehat\lambda^{VCV},v)$ in Section \ref{sec: lasso with cross-validation}. The idea here is that even though full sample estimators with the penalty parameter selected by cross-validation are hard to analyze, as discussed in Section \ref{sec: lasso with cross-validation}, the corresponding subsample estimators yield the estimators of $m_1(X_i^{\top}\beta,Y_i)$ whose consistency is relatively easy to derive. \cite{CS21} proved that the $\ell_1$-penalized estimator \eqref{eq: penalized m-estimator} with the penalty parameter selected by the multiplier bootstrap analogous to that described in Section \eqref{sec: bootstrap lasso} with preliminary estimators of $m_1(X_i^{\top}\beta,Y_i)$ obtained via subsample versions of the estimator \eqref{eq: penalized m-estimator} with the penalty parameter selected by cross-validation indeed satisfies a result of the form \eqref{eq: lasso rate of convergence}.

Finally, I note that the result of \cite{CS21} can be extended to work with time series data as well: it suffices to replace cross-validation and bootstrap by large-block-small-block versions of both cross-validation and bootstrap. Moreover, this extension is useful not only for the generalized linear models \eqref{eq: generalized linear model} but for the linear model \eqref{eq: linear model} as well. Indeed, it follows from the same arguments as those in Section \ref{sec: clustered data} that the choice of $\lambda$ in \eqref{eq: conservative lambda} may not yield a consistent Lasso estimator in the case of time series data. In fact, as argued in \cite{KPS24}, it may be rather difficult to provide a good choice of the penalty parameter $\lambda$ for the Lasso estimator in the case of time series data because the choice should depend on a measure of time-series dependence in the data. This complication can be avoided by using the large-block-small-block version of the method of \cite{CS21}.



\bibliographystyle{econ-econometrica}
\bibliography{ref}

\end{document}